\documentclass[12pt,a4paper]{article} 
\usepackage[utf8]{inputenc}
\usepackage[T1]{fontenc}
\usepackage[ngerman,english]{babel}
\usepackage{lmodern}
\usepackage[a4paper,inner=2.5cm,bottom=2.5cm,outer=2.5cm,top=2.5cm]{geometry}
\usepackage{setspace} 
\usepackage[]{graphicx}
\usepackage{epstopdf}       
\usepackage{csquotes}
\usepackage[style=authoryear,backend=bibtex8,sorting=nyt,doi=false,isbn=false,url=true,abbreviate=true,eprint=false,natbib,maxbibnames=10]{biblatex}
\usepackage{xcolor}
\usepackage{rotating}
\usepackage{float}
\usepackage{tabularx,calc}
\usepackage{amsmath,amsfonts,amssymb,amsthm}
\usepackage{bm} 
\usepackage{bbm}
\usepackage{xspace} 
\usepackage{caption}
\usepackage{subcaption}
\usepackage{multirow,array,varwidth}
\usepackage[hidelinks]{hyperref}
\usepackage{enumitem}
\usepackage{xr}
\usepackage{minitoc}
\usepackage{booktabs}
\usepackage{pifont}
\usepackage[colorinlistoftodos]{todonotes}
\usepackage{siunitx}
\usepackage{array}
\usepackage{mathtools}
\newcolumntype{H}{>{\setbox0=\hbox\bgroup}c<{\egroup}@{}} 
\usepackage{tcolorbox}
\usepackage[section]{placeins}

\AtEveryBibitem{\clearfield{note}}
\AtEveryBibitem{\clearfield{month}}
\AtEveryBibitem{\clearfield{urlyear}}
\AtEveryBibitem{\clearlist{language}}
\renewbibmacro{in:}{%
  \ifentrytype{article}{}{\printtext{\bibstring{in}\intitlepunct}}
}
\AtEveryBibitem{%
\ifentrytype{article}{\clearfield{url}}%
}

\newfloat{Algorithm}{t}{lop}



\newcommand\notsotiny{\@setfontsize\notsotiny\@vipt\@viipt}

\input{partialinput}

\begin{document}

\author{%
  Achim Ahrens\thanks{Corresponding author. ETH Z\"urich, Leonhardshalde 21, 8092 Zürich, Switzerland. {\it Email:} \url{achim.ahrens@gess.ethz.ch}}
  \and
  Christian B. Hansen\thanks{University of Chicago, United States. {\it Email:} \url{Christian.Hansen@chicagobooth.edu} (Hansen), \url{wiemann@uchicago.edu} (Wiemann).}
  \and
  Mark E. Schaffer\thanks{Heriot-Watt University, Edinburgh, United Kingdom and IZA Institute of Labor Economics. {\it Email:} \url{M.E.Schaffer@hw.ac.uk}.}
  \and
  Thomas Wiemann\footnotemark[3]
}
\title{Model Averaging and Double Machine Learning\thanks{\emph{Acknowledgment:} Many thanks to Elliott Ash, Daniel Bj\"orkegren, David Cai, Ben Jann, Michael Knaus, Rafael Lalive, Moritz Marbach, Martin Huber, Blaise Melly, Gabriel Okasa, and Martin Spindler for helpful discussions and comments. We are also thankful for the helpful feedback we have received at the AI+Economics Workshop at the ETH Z\"urich in 2022, the Italian and Swiss Stata meetings in 2022, the 2022 Machine Learning in Economics Summer Institute in Chicago, the LISER workshop ``Machine Learning in Program Evaluation, High-dimensionality and Visualization Techniques,'' the 2022 Scotland and Northern England Workshop in Applied Microeconomics, the IAAE Annual Conference in 2023, the London 2023 Stata meeting, the 2023 Stata Economics Virtual Symposium and the European Summer Meetings of the Econometric Society in 2023. We also thank anonymous reviewers for their feedback and suggestions. All remaining errors are our own. \emph{Note:} An earlier version of the paper was presented under the title ``A Practitioners' Guide to Double Machine Learning.'' \emph{Conflict of interest:} The authors declare that they have no conflict of interest. \emph{Data:} The authors provide replication code through the Journal of Applied Econometrics Data Archive and share data for all examples with the exception of the application in Section 5.1.}}

\maketitle
\thispagestyle{empty}

\begin{abstract}\singlespacing\footnotesize

This paper discusses pairing double/debiased machine learning (DDML) with \emph{stacking}, a model averaging method for combining multiple candidate learners, to estimate structural parameters. In addition to conventional stacking, we consider two stacking variants available for DDML: \emph{short-stacking} exploits the cross-fitting step of DDML to substantially reduce the computational burden and \emph{pooled stacking} enforces common stacking weights over cross-fitting folds. Using calibrated simulation studies and two applications estimating gender gaps in citations and wages, we show that DDML with stacking is more robust to partially unknown functional forms than common alternative approaches based on single pre-selected learners. We provide Stata and \textsf{R} software implementing our proposals.

\smallskip

\end{abstract}

\newpage
\setcounter{page}{1}

\renewcommand{\bm}[1]{#1}

\section{Introduction}

Motivated by their robustness to partially unknown functional forms, supervised machine learning estimators are increasingly leveraged for causal inference. For example, lasso-based approaches such as the post-double-selection lasso (PDS lasso) of \citet{Belloni2014a} have become popular estimators of causal effects under conditional unconfoundedness in applied economics \citep[e.g.][]{gilchrist2016something,dhar2022reshaping}. Yet, a recent literature also raises practical concerns about the use of machine learning for causal inference. \citet{wuthrich2021} find that lasso often fails to select relevant confounders in small samples while inference based on linear regression performs relatively well. \citet{giannone2021economic} and \citet{kolesar2023} argue that the sparsity assumption, on which the lasso fundamentally relies, is frequently not plausible in economic data sets. \citet{angrist2022} show that conditioning on confounders using random forests may yield spurious results in IV regressions.\footnote{See also \citet{Angrist2022b} for additional discussion.} In an application to the evaluation of active labor market programs, \citet{Goller2020} find that random forests are not suitable for the estimation of propensity scores. A key characteristic shared by many of these studies using machine learning for causal inference is the focus on a single pre-selected machine learner.  
 
This paper revisits the application of machine learning for causal inference in light of this recent literature. In particular, we highlight the benefits of pairing double/debiased machine learning (DDML) estimators of \citet{Chernozhukov2018} with stacking \citep{wolpert1996, Breiman1996a, laan2007}. DDML can leverage generic machine learners meeting mild convergence rate requirements for the estimation of common (causal) parameters. Stacking is a form of model averaging that allows selecting among or combining multiple candidate machine learners relying on different regularization assumptions rather than requiring an {\it ad hoc} choice between them. Based on a diverse set of applications and calibrated simulation studies, we show that the synthesis of stacking and DDML improves the robustness of estimates of target parameters to the underlying structure of the data, and illustrate the finite sample performance of stacking-based DDML estimators. The results suggest that stacking with a rich set of candidate estimators can address some of the shortcomings highlighted in the recent literature on causal inference with single pre-selected machine learners.

We further consider two alternate ways of combining stacking and DDML aimed at improving practical feasibility and stability in finite samples: {\it Short-stacking} leverages the cross-fitting step of DDML to reduce the computational burden of stacking substantially. {\it Pooled stacking} decreases the variance of stacking-based learners across the DDML cross-fitting folds. Both approaches facilitate interpretability compared to conventional stacking by enforcing common stacking weights. We complement the paper with software packages for Stata and \textsf{R} that implement the proposed approaches \citep{Ahrens2023_ddml,ddml_R}.

Model averaging techniques have a long tradition in economics and statistics. In the time-series literature, the idea of using `optimal' weights to combine forecasts goes back to the 1960s \citep{CraneCrotty1967,BatesGranger1969}. Loss-minimizing combinations of a pre-specified set of estimators were introduced under the term stacking to the statistics literature by \citet{Wolpert1992} and \citet{Breiman1996a} and generalized by \citet{laan2007}. Stacking fits a final (typically parametric) learner on a set of cross-validated predicted values derived from distinct candidate learners. A popular choice is to constrain the weights attached to each candidate learner to be non-negative and sum to one. The benefits of combining multiple estimators into a `super learner' via stacking to improve robustness to the structure of the underlying data-generating process are well-known in the econometrics and statistics literature. Under appropriate restrictions on the data generating process and loss-function, \citet{laan2003} show asymptotic equivalence between stacking and the best-performing candidate learner.\footnote{\citet{Hastie2009} and \citet{van2011targeted} provide textbook treatments of stacking and super learning. See also \citet{bhansen2012jma} for discussion of jackknife (leave-one-out) stacking. \citet{clydec2013} and \citet{le2017} provide a Bayesian interpretation of stacking in the setting where the true model is not among the candidates.} 

\begin{figure}
    \centering\scriptsize 
    \caption{Estimation bias of DDML with cross-validated lasso, feed-forward neural net and stacking}
    \label{fig:simple_example}
    \begin{subfigure}{.49\linewidth}
    \includegraphics[width=\linewidth]{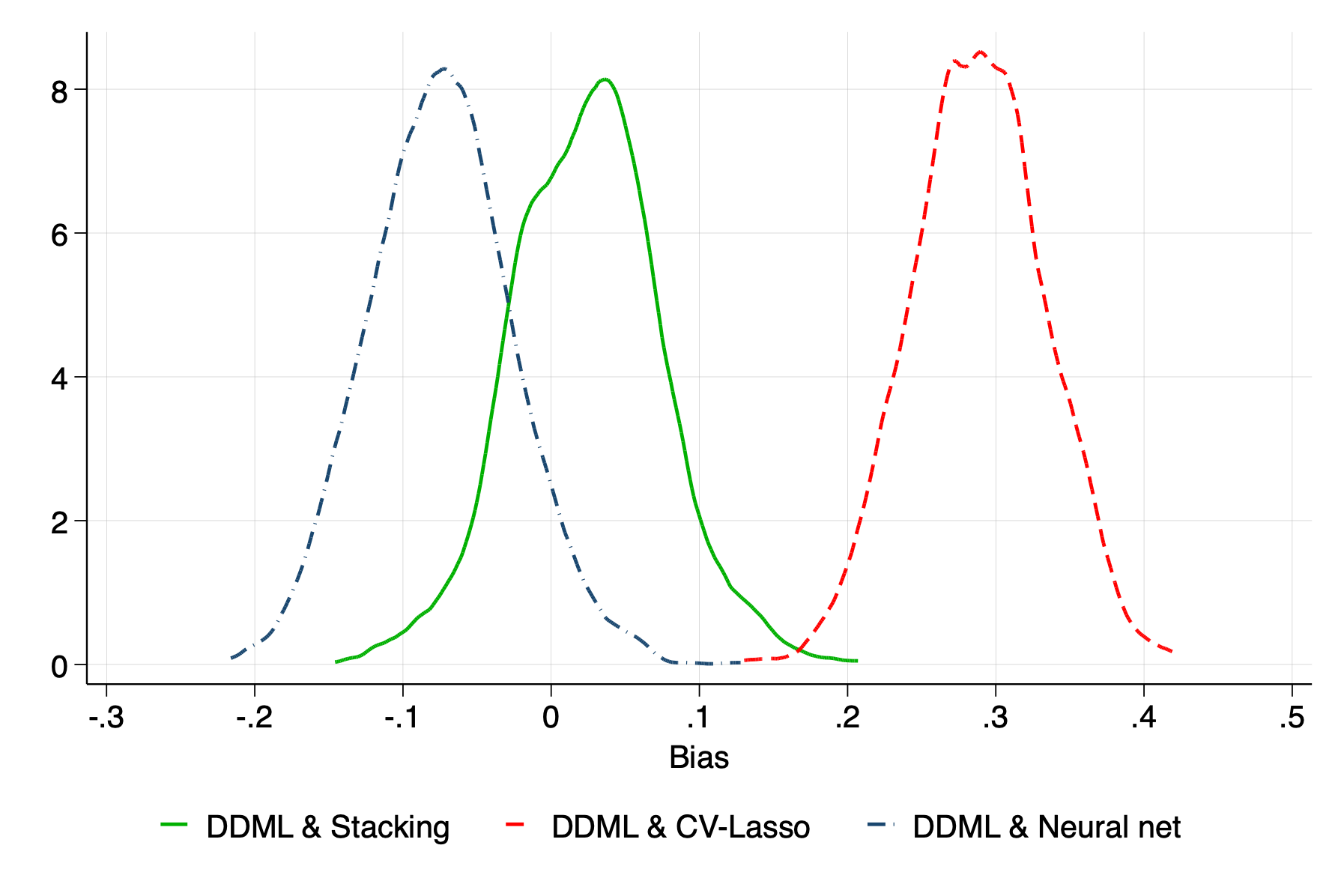}
    \caption{Non-linear DGP}
    \end{subfigure}
    \begin{subfigure}{.49\linewidth}
    \includegraphics[width=\linewidth]{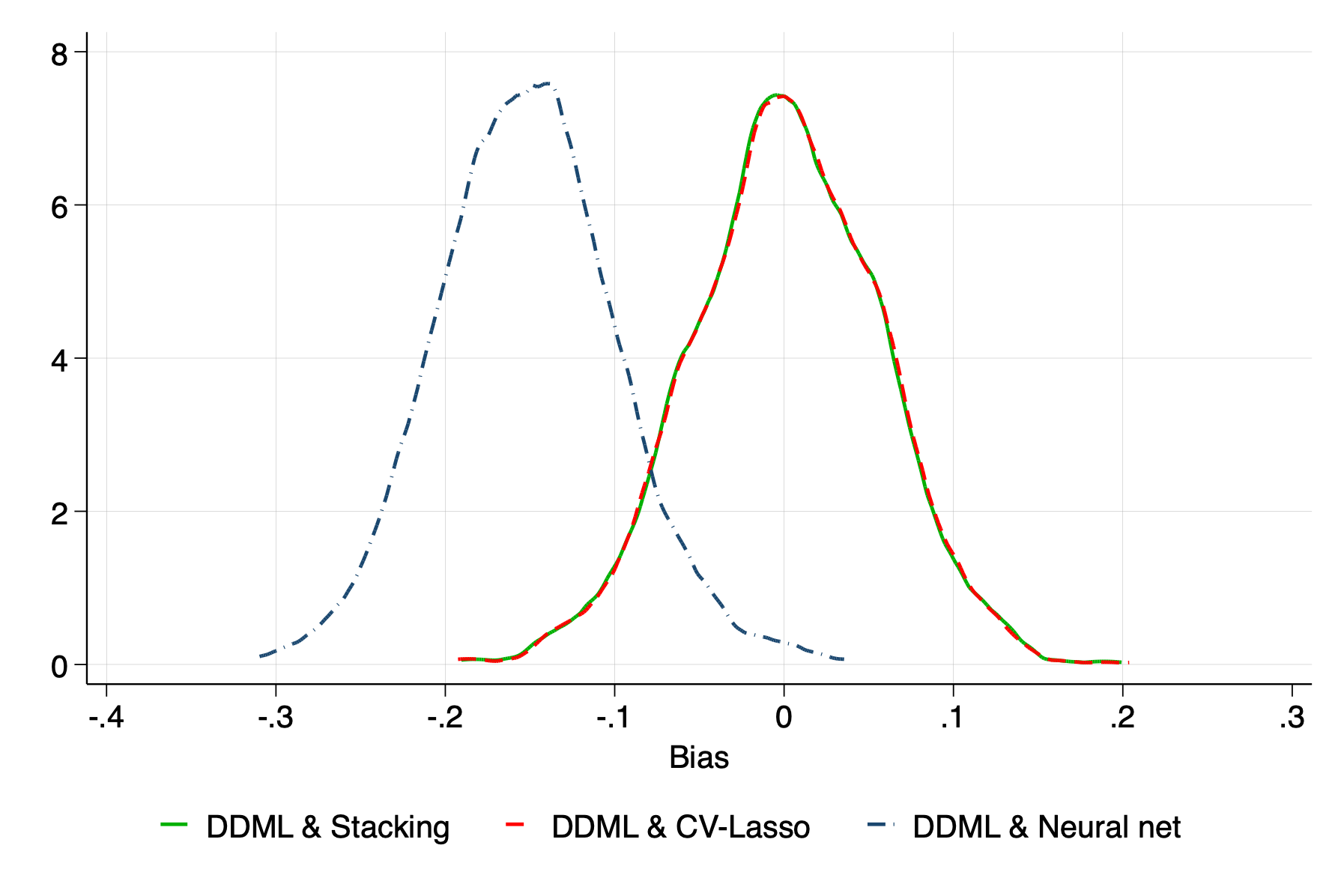}
    \caption{Approximately sparse linear DGP}
    \end{subfigure}  
    \par\medskip
    \parbox{\linewidth}{\emph{Notes:} The figures show kernel density plots comparing the bias of DDML paired with either cross-validated lasso, a feed-forward neural net (with two hidden layers of size 20) or a stacking learner combining 13 candidate learners (including cross-validated lasso and ridge, random forests, gradient-boosted trees and feed-forward neural nets). See \citet{Ahrens2023_ddml}, where this example is taken from, for details on the specification of each learner. With respect to the data-generating processes, we generate 1000 samples of size $n=1000$ using the PLM $Y_i = \theta_0D_i + c_Yg(\bm{X}_i) + \varepsilon_i$,  $D_i = c_Dg(\bm{X}_i) + u_i$ where $X_i$ are drawn from $\mathcal{N}(0,\Sigma)$ with $\Sigma_{i,k}=0.5^{\vert j-k
    \vert}$. $\varepsilon_i$ and $u_i$ are drawn from standard normal distributions.  In Figure~(a), the nuisance function is $g(\bm{X}_i)= X_{i,1}X_{i,2} + X_{i,3}^2 + X_{i,5}X_{i,5} + X_{i,6}X_{i,7}+ X_{i,8}X_{i,9} + X_{i,10}  + xX_{i,11}^2 + X_{i,12}X_{i,13}$. In Figure~(b), the nuisance function is $g(\bm{X}_i)= \sum_j 0.9^jX_{ij}$. $c_Y$ and $c_D$ are two constants chosen to ensure that the $R^2$ of the regression of $Y$ onto $X$ is approximately 0.5.}
\end{figure}

Model averaging methods and stacking are widely used in time-series forecasting and macro-econometrics \citep[for recent reviews, see][]{steel2020,Wang2023review}. Yet, despite its theoretical appeal, stacking has hitherto been rarely used for the estimation of causal effects in economics or other social sciences. An important exception is \citet{van2011targeted}, who recommend stacking in the context of Targeted Maximum Likelihood. Instead, estimators are often based on parametric (frequently linear) specifications or single pre-selected machine learners. This can have severe consequences for the properties of causal effect estimators if the given choice is ill-suited for the application at hand. A simple example is shown in Figure~\ref{fig:simple_example} which compares the performance of DDML using either cross-validated (CV) lasso or a feed-forward neural network to estimate a partially linear model across two different data-generating processes. The results show that the bias associated with each learner strongly depends on the structure of the data. Since true functional forms are often unknown in the social sciences, indiscriminate choices of machine learners in practice can thus result in poor estimates. DDML with stacking is a practical solution to this problem. As the example showcases, DDML using stacking is associated with low bias when considering a rich set of candidate learners that are individually most suitable to different structures of the data.

We conduct simulation studies calibrated to real economic datasets to demonstrate that stacking approaches can safeguard against ill-chosen or poorly tuned estimators in practical settings. Throughout, stacking estimators are associated with relatively low bias regardless of the simulated data-generating process, strongly contrasting the data-dependent performance of the causal effect estimators based on single pre-selected learners. The proposed stacking approaches thus appear relevant in the ubiquitous scenario where there is uncertainty about the set of control variables, correct functional form or the appropriate regularization assumption. 

By revisiting the simulation design of \citet{wuthrich2021}, we further show that stacking can outperform linear regression for even small sample sizes. We argue that the poor small sample performance of lasso-based approaches is partially driven by the choice of covariate transformations and illustrate how stacking can accommodate a richer set of specifications, including competing parametric models. We also find that short-stacking and pooled stacking may outperform DDML paired with conventional stacking in small to moderate sample sizes. Paired with its lower computational cost, this finding suggests that short-stacking may be an attractive baseline approach to select and combine competing reduced form specifications.

Finally, we demonstrate the value of pairing of DDML with stacking with two applications. First, we examine gender gaps in citations of articles published in top-30 economic journals from 1983 to 2020, and assess how the difference in citations change when conditioning on content and quality proxied by the abstract text. Estimating these conditional differences is a challenging statistical problem due to the non-standard nature of text data, which is increasingly encountered in economic applications \citep[see also e.g.,][]{ash2023a,chen2023gender,eberhardt2022}. Second, we revisit a UK sample of the OECD Skill Survey to estimate semiparametric Kitagawa-Oxaca-Binder estimates of the unexplained gender wage gap. Both applications highlight that estimators of structural parameters based on single learners can be highly sensitive to the underlying structure of the data and/or poor tuning. The applications further demonstrate that DDML with stacking is a simple and practical solution to resolve the difficult problem of choosing a particular candidate learner in practice. Further, we observe that the optimal stacking weights often vary across reduced-form equations --- meaning that different conditional expectation functions in the same data set are best estimated using different learners. This behavior sharply contrasts with common estimation approaches, such as OLS and PDS lasso, that impose the same form for each conditional expectation function.

The remainder of the paper is organized as follows: Section \ref{sec:theory} provides a brief review of DDML. Section~\ref{sec:stacking} discusses DDML with stacking, short-stacking, and pooled stacking. Section \ref{sec:simulation} presents our calibrated simulation studies. Section~\ref{sec:applications} discusses the applications, and Section~\ref{sec:conclusion} concludes.

\section{Double/Debiased Machine Learning}\label{sec:theory}

This section outlines double/debiased machine learning as discussed in \citet{Chernozhukov2018}. Throughout, we focus on the partially linear model as a natural extension of commonly applied linear regression methods. Despite its simplicity, the partially linear model illustrates practical challenges in the application of DDML that can be addressed by stacking. We highlight, however, that our discussion also applies to the wide range of models outlined in \citet{Chernozhukov2018} and more generally to estimation of low-dimensional structural parameters in the presence of high-dimensional nuisance functions.\footnote{A key example not explicitly discussed in \citet{Chernozhukov2018} is doubly-robust estimation of difference-in-difference parameters with staggered treatment assignment as in \citet{callaway2021difference} and \citet{chang2020double}. In settings with conditional parallel trends assumptions, high-dimensional nuisance functions arise in the estimation of group-time specific average treatment effect on the treated. The pairing of DDML and stacking, as proposed in this paper, also directly applies to the estimator of \citet{callaway2021difference} under a conditional unconfoundedness assumption.} Stacking could also be applied in settings where DDML is applied with nonparametric targets as in \citet{ColangeloLee}.

The partially linear model is defined by a random vector $(Y, D, X^\top, U)$ with joint distribution characterized by \begin{align}
    Y = \theta_0D + g_0(X) + U, \label{eq:partial_linear_model}
\end{align}
where $Y$ is the outcome, $D$ is the scalar variable of interest, and $X$ is a vector of control variables. The parameter of interest $\theta_0$ and the unknown nuisance function $g_0$ are such that the corresponding residual $U$ satisfies the conditional orthogonality property $E[Cov(U, D\vert X)]= 0$. These properties are analogous to the orthogonality properties of residuals in multiple linear regression with the key difference here being that $g_0$ need not be linear in the controls. 

Albeit a seemingly small change in specification, the partially linear model has several important advantages over linear regression. For discrete $D$, for example, results in \citet{angrist1999empirical} imply that $\theta_0$ can be interpreted as a positively weighted average of incremental changes in the conditional expectation function $E[Y\vert D = d, X]$. Under appropriate conditional unconfoundedness assumptions, $\theta_0$ thus corresponds to a convex combination of conditional average treatment effects.\footnote{Similarly, for continuous $D$, $\theta_0$ corresponds to a positively weighted average of derivatives of the conditional expectation function $E[Y\vert D = d, X]$ with respect to $d$. Under a conditional unconfoundedness assumption, $\theta_0$ is thus a convex combination of derivatives of the causal response function.} Importantly, these interpretations remain valid even if the additive separability assumption of the partially linear model fails. Linear regression coefficients, in contrast, do not correspond to positively weighted averages of causal effects without imposing strong linearity assumptions that are questionable in real applications.\footnote{In the context of IV estimation where instrument validity relies on observed confounders, \citet{blandhol2022tsls} emphasize that, in the absence of strong functional form assumptions, two stage least squares does not generally correspond to a convex combination of local average treatment effects (LATE). We note that the IV analogue to the partially linear model does admit a causal interpretation under the LATE assumptions, just as the partially linear model admits a weakly causal interpretation under conditional unconfoundedness.}

The advantages of the partially linear model in the interpretation of its parameter of interest come at the cost of a more challenging estimation problem relative to estimating a model that is linear in a pre-specified set of variables. Estimators for $\theta_0$ are based on the solution to the moment equation \begin{align*}
    E\left[\left(Y - \ell_0(X) - \theta_0(D - m_0(X))\right)\left(D - m_0(X)\right)\right] = 0,
\end{align*}
given by
\begin{align*}
    \theta_0 = \frac{E\left[\big(Y - \ell_0(\bm{X})\big)\big(D - m_0(\bm{X})\big)\right]}{E\left[(D - m_0(\bm{X}))^2\right]},
\end{align*}
where $\ell_0(X) \equiv E[Y \vert X]$ and $m_0(X) \equiv E[D \vert X]$ are the conditional expectations of the outcome and variable of interest given the controls, respectively. Since conditional expectation functions are high-dimensional in the absence of strong functional form assumptions, a sample analogue estimator for $\theta_0$ requires nonparametric first-step estimators for the nuisance parameters $\ell_0$ and $m_0$. While nonparametric estimation generally reduces bias compared to linear regression alternatives, the increased variance associated with more flexible functional form estimation introduces additional statistical challenges: To allow for statistical inference on $\theta_0$, the nonparametric estimators need to converge sufficiently quickly to the true conditional expectation functions as the sample size increases.

DDML defines a class of estimators that allows for statistical inference on the parameter of interest $\theta_0$ while only imposing relatively mild convergence requirements on the nonparametric estimators. These mild requirements are central to the wide applicability of DDML as they permit the use of a large variety of machine learners.\footnote{The exact convergence rate requirement for nonparametric estimators depends on the parameter of interest. \citet{Chernozhukov2018} name the crude rate requirement of $o(n^{-1/4})$, but provide examples where the rate requirement is considerably weaker. Recent contributions show that these requirements are satisfied by specific instances of machine learners; see, e.g., results for lasso \citep{bickel2009simultaneous, Belloni2012}, random forests \citep{wager2015adaptive, wager2018, Athey2019a}, neural networks \citep{schmidt2020nonparametric, farrell2021deep}, and boosting \citep{luo2016high}. The exact asymptotic properties of many other machine learners remain an active research area.} 

Two key devices permit the mild convergence requirements of DDML: Identification of the parameter of interest based on Neyman-orthogonal moment conditions and estimation using cross-fitting. Neyman-orthogonal moment conditions are insensitive to local perturbations around the true nuisance parameter.\footnote{In the context of the partially linear model, the formal Neyman-orthogonality requirement is \begin{align*}
        0=&\frac{\partial}{\partial \lambda}E\Big[\big(Y - \{\ell_0(X) + \lambda(\ell(X) - \ell_0(X))\} - \tau_0(D- \{m_0(X) + \lambda(m(X) - m_0(X))\})\big) \\
        &\qquad \qquad  \times \big(D- \{m_0(X) + \lambda(m(X) - m_0(X))\}\big)\Big]\bigg\vert_{\lambda=0}
\end{align*}
for arbitrary measurable functions $\ell$ and $m$, which can easily be verified using properties of the residuals.} Cross-fitting is a sample-splitting approach that addresses the {\it own-observation bias} that arises when the nuisance parameter estimation and the estimation of $\theta_0$ are applied to the same observation. In practice, cross-fitting is implemented by randomly splitting a sample $\{(Y_i, D_i, X^\top_i)\}_{i\in I}$ indexed by $I = \{1, \ldots, n\}$ into $K$ evenly-sized folds, denoted as $I_1,\ldots, I_K$. For each fold $k$, the conditional expectations $\ell_0$ and $m_0$ are estimated using only observations not in the $k$th fold --- i.e., in $I^c_k\equiv I \setminus I_k$ --- resulting in $\hat{\ell}_{I^c_{k}}$ and $\hat{m}_{I^c_{k}}$, respectively, where the subscript ${I^c_{k}}$ indicates the subsample used for estimation. The out-of-sample predictions for an observation $i$ in the $k$th fold are then computed via $\hat{\ell}_{I^c_{k}}(\bm{X}_i)$ and $\hat{m}_{I^c_{k}}(\bm{X}_i)$. Repeating this procedure for all $K$ folds then allows for computation of the DDML estimator for $\theta_0$:\begin{align*}
        \hat{\theta}_n = \frac{\frac{1}{n}\sum_{i=1}^n \big(Y_i-\hat{\ell}_{I^c_{k_i}}(X_i)\big)\big(D_i-\hat{m}_{I^c_{k_i}}(X_i)\big)}{\frac{1}{n}\sum_{i=i}^n \big(D_i-\hat{m}_{I^c_{k_i}}(X_i)\big)^2},
\end{align*}
where $k_i$ denotes the fold of the $i$th observation. 

Since the cross-fitting algorithm depends on the randomized fold split, and since some machine learners rely on randomization too, DDML estimates vary with the underlying random-number generator and seed. To reduce dependence on randomization, it is thus worthwhile to repeat the cross-fitting procedure and apply mean or median aggregation over DDML estimates \citep[see Remark~2 in][]{Ahrens2023_ddml}. We show in Section~\ref{sec:applications} that repeating the cross-fitting procedure is a useful diagnostic tool, allowing to gauge the stability of DDML estimators.

Under the conditions of \citet{Chernozhukov2018} --- including, in particular, the convergence requirements on the nonparametric estimators --- $\hat{\theta}_n$ is root-$n$ asymptotically normal around $\theta_0$. As already highlighted by the example in Figure~\ref{fig:simple_example}, however, a poorly chosen or poorly tuned machine learner for the estimation of nuisance parameters $\hat{\ell}$ and $\hat{m}$ can have detrimental effects on the properties of $\hat{\theta}_n$. Since no machine learner can be best across all settings, this raises the difficult question of which learner to apply in a particular setting. In the next section, we discuss how DDML can be paired with stacking to provide a practical solution to the choice of learner. We also illustrate how the cross-fitting structure naturally arising in DDML estimators can be leveraged to substantially reduce the computational burden otherwise associated with stacking.

\section{Pairing DDML with Stacking Approaches}\label{sec:stacking}

This section discusses the estimation of structural parameters by pairing DDML with stacking approaches. After the discussion of DDML with conventional stacking, we introduce two stacking variants that leverage the cross-fitting structure of DDML estimators: short-stacking and pooled stacking. To fix ideas, we focus on the nuisance parameter $\ell_0(X)=E[Y\vert X]$ arising in the partially linear model where we consider an i.i.d.\ sample $\{(Y_i,X_i)\}_{i\in I}$. Further, we consider a rich set of $J$ pre-selected base or candidate learners. The set of learners could include distinct parametric and nonparametric estimators --- e.g., linear or logistic regression, regularized regression such as the lasso, or tree-based methods such as random forests --- as well as the same algorithm with varying (hyper-) tuning parameters or different (basis) expansions of the control variables.
It is important to note that the set of candidate learners for stacking can readily incorporate commonly used unregularized learners such as linear or logistic regression; in practice, sometimes the best-performing candidate learner may be one such learner.

\begin{figure}[htbp]
    \centering\scriptsize
    \caption{Cross-fitting with conventional stacking}
    \includegraphics[width=.8\linewidth]{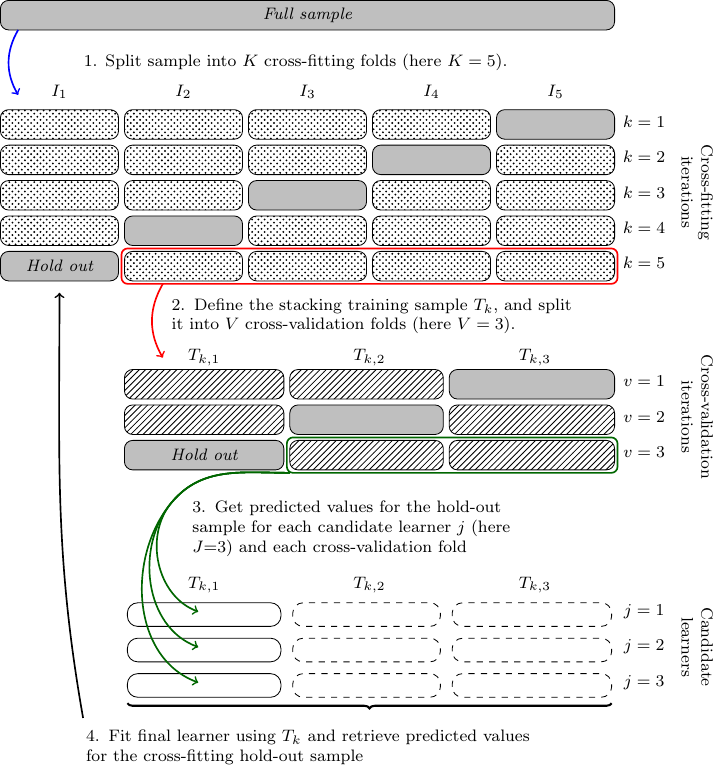}
    \par\medskip
    \parbox{\linewidth}{\emph{Notes:} The diagram illustrates cross-fitting with conventional stacking. The diagram uses $K=5$ cross-fitting folds, $V=3$ cross-validation folds and $J=3$ candidate learners. Step 1: The sample randomly is split into cross-fitting folds $I_1,\ldots,I_K$. Step 2: In each step $k\in \{1,\ldots,K\}$ of the cross-fitting process, we define the training data as $T_k\equiv I\setminus I_k$. The $k$-step training data is then split into $V$ sub-partitions, which we denote by $T_{k,1}, \ldots, T_{k,V}$. Step 3: For each cross-fit step $k$ and cross-validation step $v\in \{1,\ldots,V\}$, fit each base learner $j\in \{1,\ldots,J\}$ on $T_k\setminus T_{k,v}$ and obtain predicted values for the cross-validation hold-out sample. Step 4: Fit the final learner on sample $T_k$ to obtain predicted values for the cross-fitting hold-out sample $I_k$.}
    \label{fig:stacking_diagram}
\end{figure}

\paragraph{DDML with conventional stacking.} Combining DDML with conventional stacking involves two layers of re-sampling, as we illustrate in Figure~\ref{fig:stacking_diagram}. The {\it cross-fitting layer} divides the sample into $K$ cross-fitting folds, denoted by $I_1,\ldots,I_K$. In each cross-fitting step $k\in\{1,\ldots,K\}$, the stacking learner is trained on the training sample which excludes fold $I_k$ and which we label $T_k\equiv I\setminus I_k$. Fitting the stacking learner, in turn, requires sub-dividing the training sample $T_k$ further into $V$ cross-validation folds. This second sample split constitutes the {\it cross-validation layer}. We denote the cross-validation folds in cross-fitting step $k$ by $T_{k,1},\ldots,T_{k,V}$. Each candidate learner $j\in\{1,\ldots,J\}$ is cross-validated on these folds, yielding cross-validated predicted values for each learner. 

The final learner fits the outcome $Y_i$ against the cross-validated predicted values of each candidate learner. The most common choice is to construct a convex combination via constrained least squares (CLS), with weights restricted to be non-negative and summing to one. Specifically, for each $k$, candidate learners are combined to solve 
\[ \underset{w_{k,1},\ldots,w_{k,J}}{\min} \sum_{i\in T_k} \left( Y_i - \sum_{j=1}^J w_{k,j} \hat{\ell}^{(j)}_{T_{k,v(i)}^c}(\bm{X}_i)  \right)^2   \qquad \textrm{s.t.}\ w_{k,j}\geq 0,\ \sum_{j=1}^J |w_{k,j}|=1.\]
Here, $\hat{\ell}^{(j)}_{T_{k,v(i)}^c}(\bm{X}_i)$ denotes the out-of-sample predicted value for observation $i$, which is calculated from training candidate learner $j$ on sub-sample $T_{k,v(i)}^c \equiv T_k \setminus T_{k,v(i)}$, i.e., all step-$k$ cross-validation folds but fold $(k,v(i))$ which is the fold of the $i$th observation. We call the resulting $\hat{w}_{k,j}$ the {\it stacking weights}. The stacking predictions are obtained as $\sum_j \hat{w}_{k,j} \hat{\ell}_{T_k}^{(j)}(\bm{X}_i)$ where each learner $j$ is re-fit on $T_k$. 

Although various options for combining candidate learners are available, CLS facilitates the interpretation of stacking as a weighted average of candidate learners \citep{Hastie2009}. Due to this constraint, CLS tends to set some stacking weights to exactly zero. The constraint also regularizes the final estimator, which is important to mitigate issues arising from potential multicollinearity of the candidate learners. An alternative to CLS, which we refer to as {\it single-best learner}, is to impose the constraint that $w_{k,j}\in \{0,1\}$ and $\sum_{j} w_{k,j}=1$, implying that only the candidate learner with lowest cross-validated loss is used as the final estimator. Under appropriate restrictions on the data-generating process and loss function, \citet{laan2003} show asymptotic equivalence between stacking and the best-performing candidate learner.\footnote{The {\it scikit-learn} \citep{sklearn_api} routines {\tt StackingRegressor} and {\tt StackingClassifier} implement stacking for Python. In Stata, stacking regression and classification are available via \texttt{pystacked}, which is a Stata front-end for these Python routines \citep{Ahrens2022}.} 

A drawback of DDML with stacking is its computational complexity. Considering the estimation of a single candidate learner as the unit of complexity (and ignoring the cost of fitting the final learner), DDML with stacking heuristically has a computational cost proportional to $K\times V \times J$. For example, when considering DDML with $K=5$ cross-fitting folds and $J=10$ candidate learners that are combined based on $V=5$ fold cross-validation, more than 250 candidate learners need to be individually estimated. Although DDML with stacking is ``embarrassingly parallel'' and can thus be expected to decrease in computational time nearly linearly in the number of available computing processes, the increased complexity limits its application to moderately complex applications. Another potential concern (which we investigate in Section~\ref{sec:simulation_very_small}) is that DDML with stacking might not perform well in small samples, given that candidate learners are effectively trained on approximately $\frac{(K-1)(V-1)}{KV}$\% of the full sample (see Figure~\ref{fig:stacking_diagram}). These two concerns motivate \emph{short-stacking}.

\begin{figure}[htbp]
    \centering\scriptsize
    \caption{Cross-fitting with \emph{short-}stacking}
    \includegraphics[width=.8\linewidth]{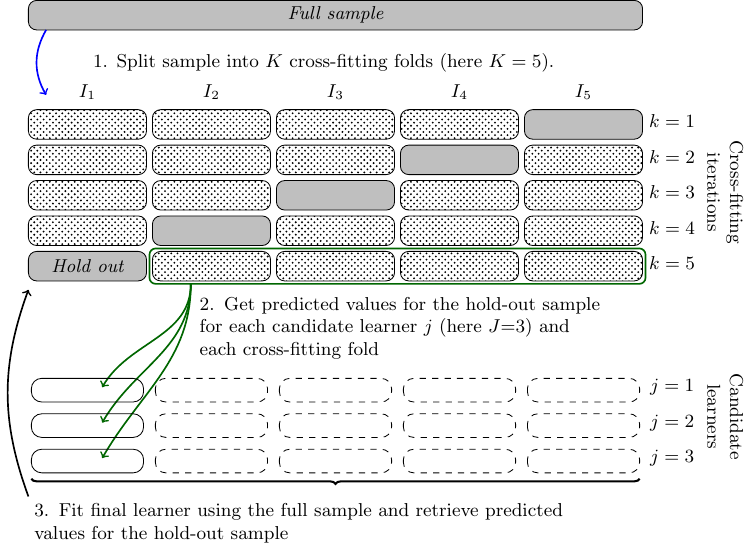}
    \par\medskip
    \parbox{\linewidth}{\emph{Notes:} The diagram illustrates cross-fitting with short-stacking. The diagram uses $K=5$ cross-fitting folds and $J=3$ candidate learners. Step 1: The sample randomly is split into cross-fitting folds $I_1,\ldots,I_K$. Step 2: In each step $k\in \{1,\ldots,K\}$ of the cross-fitting process, we define the training data as $I^c_k=I\setminus I_k$. Fit each base learner $j\in \{1,\ldots,J\}$ on the training data and obtain predicted values for the cross-fitting hold-out sample $I_k$. Step 3: Fit the final learner on the full sample to obtain predicted values for the cross-fitting hold-out sample.}
    \label{fig:short_diagram}
\end{figure}

\paragraph{DDML with short-stacking.} 
In the context of DDML, we propose to take a short-cut: Instead of fitting the final learner on the cross-{\it validated} fitted values in each step $k$ of the cross-fitting process, we can directly train the final learner on the cross-{\it fitted} values using the full sample; see Figure~\ref{fig:short_diagram}. Formally, candidate learners are combined to solve 
\[  \underset{w_1,\ldots,w_J}{\min} \sum_{i=1}^n \left( Y_i - \sum_{j=1}^J w_j\hat{\ell}^{(j)}_{I_{k(i)}^c}(\bm{X}_i) \right)^2  \qquad \textrm{s.t.}\ w_j\geq 0,\ \sum_j |w_j|=1\]
where $w_j$ are the short-stacking weights. Cross-fitting thus serves a double purpose: First, it avoids the own-observation bias by avoiding overlap between the samples used for estimating high-dimensional nuisance functions and the samples used for estimating structural parameters. Second, it yields out-of-sample predicted values which we leverage for constructing the final stacking learner. As a consequence, the computational cost of DDML with short stacking is heuristically only proportional to $K\times J$ in units of estimated candidate learners. In the example from the previous paragraph, short-stacking thus requires estimating about 200 fewer candidate learners. 

We recommend DDML with short-stacking in settings where the number of candidate learners is small relative to the sample size, i.e., $J\ll n$. We believe this setting provides a good approximation to current applications of machine learning in economics and other social sciences where it is rare to consider more than a few candidate learners. If instead the number of considered learners is very large relative to the sample size --- i.e., settings in which inference for standard linear regression on $J$ variables is invalid --- pairing DDML with short-stacking may introduce bias.\footnote{Suppose, for simplicity, we consider ordinary (unconstrained) least squares as the final learner. Heuristically, the regression of $Y_i$ against $J$ sets of cross-fitted predicted values is akin to a conventional least squares regression of $Y_i$ against $J$ observed regressors where good performance would require $J/n\rightarrow 0$, ignoring that the cross-fitted predicted values are estimated. The additional regularization by constrained least squares should further weaken this rate requirement.} 

\paragraph{DDML with pooled stacking.} While DDML with conventional stacking has one vector of weights per cross-fitting fold, short-stacking yields a single weight for each learner. A single weight for each learner decreases the variance of the final estimator and facilitates the interpretation of the stacking weights. Another way of achieving common stacking weights is DDML with pooled stacking. Pooled stacking relies on the same two-layer re-sampling strategy as conventional stacking, but combines candidate learners to solve
\[ \underset{w_{1},\ldots,w_{J}}{\min} \sum_{i\in I}\sum_{k\neq k(i)} \left( Y_i - \sum_{j=1}^J w_{j} \hat{\ell}^{(j)}_{T_{k,v(i)}^c}(\bm{X}_i)  \right)^2   \qquad \textrm{s.t.}\ w_{j}\geq 0,\ \sum_{j=1}^J |w_{j}|=1.\]
That is, pooled stacking collects the cross-validated predicted values that are calculated in each step $k$ of the cross-fitting process for each learner $j$ and estimates the stacking weights based on the pooled data set. We note that the computational costs are approximately the same as for DDML with conventional stacking.

\section{The Practical Benefits of DDML with Stacking: Two Simulation Studies}\label{sec:simulation}

In this section, we discuss two simulation studies illustrating the advantages of pairing DDML with stacking over alternative approaches based on single pre-selected learners. We begin with a simulation calibrated to household data on wealth and 401k eligibility from the 1991 wave of the Survey of Income and Program Participation (SIPP) in Section \ref{sec:advantages_stacking}. In Section \ref{sec:simulation_very_small}, we revisit the simulation of \citet{wuthrich2021} to assess the robustness of DDML with stacking approaches in very small samples.

\subsection{Simulation calibrated to the SIPP 1991 household data}\label{sec:advantages_stacking}

To assess the performance of DDML with conventional stacking, short-stacking and pooled stacking in a realistic setting, we consider the analysis of 401(k) eligibility and total financial assets in \citet{poterba1995} as the basis for an empirically calibrated Monte Carlo simulation. The application has recently been revisited by \citet{Belloni2017}, \citet{Chernozhukov2018}, and \citet{wuthrich2021} to approximate high-dimensional confounding factors using machine learning. We focus on estimating the partially linear model discussed in the previous section. The outcome is measured as net financial assets, the treatment variable is an indicator for eligibility to the 401(k) pension scheme, and the set of controls includes age, income, education in years, family size, as well as indicators for two-earner status, home ownership, and participation in two alternative pension schemes.

\begin{Algorithm}
    \footnotesize\onehalfspacing\centering
    \caption{Algorithm for the calibrated Monte Carlo simulation}\label{algo:calibrated_sim}
    \begin{minipage}{.93\linewidth}
    \begin{tcolorbox}[standard jigsaw,opacityback=0]
    Let $\{(y_i, d_i, x_i)\}_{i = 1, \ldots, n}$ denote the observed sample, where $i$ is a household in the 1991 SIPP and $y_i$, $d_i$, and $x_i$ respectively denote net financial assets, an indicator for 401(k) eligibility, and the vector of control variables. 
    \begin{enumerate}[nosep] 
        \item Using the full sample, obtain the slope coefficient $\hat{\theta}_{OLS}\approx  5\,896$ from linear regression of $d_i$ against $d_i$, and $x_i$ in the original data. Construct the partial residuals $y_i^{(r)} = y_i - \hat{\theta}_{OLS} d_i, \, \forall i$.
        \item Fit a supervised learning estimator (either linear regression or gradient boosting) to predict $y_i^{(r)}$ with the controls $x_i$. Denote the fitted estimator by $\Tilde{g}$. Similarly, fit a supervised learning estimator to predict $d_i$ with $x_i$ and denote the fitted estimator by $\Tilde{h}$.
        \item Repeat to generate simulated samples of size $n_b$:
        \begin{enumerate}[nosep] 
        \item Sample from the empirical distribution of $x_i$ by bootstrapping $n_b$ observations from the original data. Denote the bootstrapped sample by $\mathcal{D}_b$. 
        \item Draw $\nu_i \overset{iid}{\sim} \mathcal{N}(0, \kappa_1)$ and $\varepsilon_i \overset{iid}{\sim} \mathcal{N}(0, \kappa_2)$, where $\kappa_1$ and $\kappa_2$ are simulation hyperparameters. Define
        \begin{align*}
            \Tilde{d}^{(b)}_i &= \mathbbm{1}\{\Tilde{h}(x_i) + \nu_i \geq 0.5\}  \\
            \Tilde{y}^{(b)}_i &= \theta_0 \Tilde{d}^{(b)}_i + \Tilde{g}(x_i) + \varepsilon_i \qquad\forall i \in \mathcal{D}_b
        \end{align*}
        where we set $\theta_0=6\,000$ to roughly resemble the magnitude of the regression coefficient of 401(k) eligibility in the full data. 
        \end{enumerate}
    \end{enumerate}
    \end{tcolorbox}\end{minipage}\par\singlespacing\scriptsize\vspace*{-.3cm}
    \parbox{\linewidth}{{\it Notes:} We set the hyper-parameter $\kappa_1$ and $\kappa_2$ to approximately match variance of 401(k) eligibility and wealth in the data. The values of the simulation hyperparameters $(\kappa_1, \kappa_2)$ differ slightly depending on the supervised learning estimator used to fit the reduced form equations in the data. We take $\kappa_1= 0.35$ in both scenarios but take $\kappa_2 = 55\,500$ when using linear regression and $\kappa_2 = 54\,000$ when using gradient boosting. Differences arise because gradient boosting reduces residual variance in the true data.}
\end{Algorithm}

The simulation involves three steps. In the calibration step, we fit two generative models to the $n=9\,915$ households from the 1991 wave of the Survey of Income and Program Participation. The first generative model is fully linear while the second is partially linear, allowing controls to enter non-linearly through gradient-boosted trees fitted to the real data. This approach is aimed at extracting and magnifying the linear or non-linear structures in the empirical conditional distributions, respectively, enabling us to compare the performance of estimators across favorable and unfavorable structures of the data. The generative step then simulates datasets of size $n_b=\{9\,915,\,99\,150\}$ from the respective fully linear model and the partially linear model. Throughout, we set the effect of 401(k) eligibility on total financial wealth to $\theta_0=6\,000$. Finally, in the estimation step, we fit various estimators to bootstrapped samples of the generated datasets and assess their statistical properties. We outline the steps used for constructing the two generative models in more detail in Algorithm~\ref{algo:calibrated_sim}.

For each bootstrap sample, we calculate estimates of the effect of 401(k) eligibility on simulated net financial assets. The estimators we consider are linear regression, the post-double selection (PDS) lasso estimator proposed by \citet{Belloni2014a}, as well as DDML estimators with and without stacking. The candidate learners of the DDML estimators are linear regression, cross-validated lasso and ridge regression with interactions and second-order polynomial expansions of the controls, cross-validated lasso and ridge with no interactions but 10\textsuperscript{th}-order polynomial expansions of the controls, two versions of random forests, two versions of gradient-boosted trees, and feed-forward neural nets with three hidden layers of size five (see Table~\ref{tab:sim_advantages_bias} notes for details). We estimate DDML paired with conventional stacking, short-stacking and pooled stacking, and consider different methods to construct the final conditional expectation function estimator: CLS, unconstrained linear regression (OLS), selecting the single best estimator, and an unweighted average.

Table~\ref{tab:sim_advantages_bias} presents the mean bias, median absolute bias (MAB) and coverage rates of a 95\% confidence interval associated with estimates of the effect of 401(k) eligibility on net financial assets. The left and right panels correspond to results based on data simulated from the linear (Panel~A) and non-linear (Panel~B) generative models, respectively. The CLS weights associated with each candidate learner are shown in Table~\ref{tab:sim_advantages_weights_cls}.\footnote{Further results are provided in the Appendix. Table~\ref{tab:sim_advantages_mspe} in the Appendix gives the mean-squared prediction errors (MSPE) for each candidate learner for comparison. Table~\ref{tab:sim_advantages_bias_se} reports standard errors of the bias. Tables~\ref{tab:sim_advantages_weights_ols} and \ref{tab:sim_advantages_weights_singlebest} show the stacking weights when using single-best and OLS as the final learner, respectively.}

    \begin{sidewaystable} 
        \centering\scriptsize\singlespacing
        \caption{Bias and Coverage Rates in the Linear and Non-Linear DGP}
        \sisetup{ 
            table-number-alignment=left,
            tight-spacing=true,
            table-format=-4.1 
        }
        \label{tab:sim_advantages_bias}
            \begin{tabular}{rlS[table-format=-4.1]S[table-format=4.1]ccS[table-format=-4.1]S[table-format=4.1]ccS[table-format=-4.1]S[table-format=4.1]ccS[table-format=-4.1]S[table-format=4.1]c}
        \toprule
        \midrule
        && \multicolumn{7}{c}{\textit{Panel (A): Linear DGP}} & &\multicolumn{7}{c}{\textit{Panel (B): Non-linear DGP}}\\
        \cmidrule{3-9}\cmidrule{11-17}
              &       & \multicolumn{3}{c}{$n_b = 9\,915$} &       & \multicolumn{3}{c}{$n_b = 99\,150$}
              &       & \multicolumn{3}{c}{$n_b = 9\,915$} &       & \multicolumn{3}{c}{$n_b = 99\,150$}\\
        \cmidrule{3-5}\cmidrule{7-9}\cmidrule{11-13}\cmidrule{15-17}
        && \multicolumn{1}{c}{Bias}  & \multicolumn{1}{c}{MAB}    & Rate  &       & \multicolumn{1}{c}{Bias}  & \multicolumn{1}{c}{MAB}   & Rate & & \multicolumn{1}{c}{Bias}  & \multicolumn{1}{c}{MAB}    & Rate  &       & \multicolumn{1}{c}{Bias}  & \multicolumn{1}{c}{MAB}   &\multicolumn{1}{c}{Rate} \\ \midrule
              \multicolumn{4}{l}{Full sample:}\\
              \partialinput{1}{2}{Simul/sim_Advantages/sim_output_folds2}
               \multicolumn{4}{l}{DDML methods:}\\
               \multicolumn{4}{l}{\it ~~Candidate learners}\\
              \partialinput{3}{12}{Simul/sim_Advantages/sim_output_folds2}
              \multicolumn{4}{l}{\it ~~Stacking approaches}\\
              \partialinput{13}{24}{Simul/sim_Advantages/sim_output_folds2}
        \midrule
        \bottomrule
        \end{tabular}%
        \par\medskip
      \parbox{\linewidth}{%
      \textit{Notes:} The table reports mean bias, median absolute bias (MAB) and coverage rate of a 95\% confidence interval for the listed estimators.
      We consider DDML with $K=2$ cross-fit folds and the following individual learners: OLS with elementary covariates, CV lasso and CV ridge with second-order polynomials and interactions, CV lasso and CV ridge with 10th-order polynomials but no interactions, random forest with low regularization (8 predictors considered at each leaf split, no limit on the number of observations per node, bootstrap sample size of 70\%), highly regularized random forest (5 predictors considered at each leaf split, at least 10 observation per node, bootstrap sample size of 70\%), gradient-boosted trees with low regularization (500 trees, maximum depth of 3 and a learning rate of 0.01), gradient-boosted trees with high regularization (250 trees, maximum depth of 3 and a learning rate of 0.01), feed-forward neural nets with three hidden layers of size five.
      For reference, we report two estimators using the full sample: OLS and PDS lasso.  Finally, we report results for DDML paired with conventional stacking, short-stacking and pooled stacking where the final estimator is either CLS, OLS, the unweighted average of candidate learners or the single-best candidate learner. Results are based on 1\,000 replications.
      }
    \end{sidewaystable}

Given the construction of the generative models, we would expect that linear regression performs best in the fully linear setting and that DDML with gradient boosting performs best in the nonlinear setting where the nuisance function is generated by gradient boosting. The simulation results confirm this intuition, showing that the two procedures achieve among the lowest bias and median absolute bias in the data-generating processes that are based on them. Researchers are rarely certain of the functional structure in economic applications, however, so that it is more interesting to consider their respective performance in the non-favorable setting. In the non-linear data-generating process, linear regression is among the estimators with the worst performance across all three measures. Similarly, gradient boosting-based DDML is non-optimal in the linear data-generating process. It is outperformed by linear regression and CV lasso, both of which enforce a linear functional form on the control variables, in terms of MAB. 

The simulation results are consequences of the ``no free lunch'' theorem in machine learning \citep{wolpert1996}. Informally, the theorem states that there exists no estimator that performs best across all empirical settings. Researchers must, therefore, carefully match estimators to their application. However, with limited knowledge about underlying data-generating processes and few functional form restrictions implied by economic theory, the number of plausibly suitable estimators is typically large. 

The bottom section of Table~\ref{tab:sim_advantages_bias} reports results for DDML combined with the three stacking approaches outlined in Section~\ref{sec:stacking}. For each stacking approach, we consider stacking weights estimated by (CLS) as outlined in Section \ref{sec:stacking}, set equal to $1/J$ (Average), estimated without constraint by OLS (OLS), and by selecting only the single best candidate learner (Single-best). We find that short-stacking performs similarly to, and sometimes better than, conventional and pooled stacking, while being computationally much cheaper (as shown in Table~\ref{tab:timing}). For example, at $K=10$ and $V=5$, DDML combined with short-stacking ran around 4.3 times faster on the full sample than DDML with conventional or pooled stacking, which is roughly in line with a speed improvement by a factor of $1/V$.\footnote{The computations were performed on the high-performance cluster of the ETH Zurich. Each instance used a single core of an AMD EPYC processor with 2.25-2.6GHz (nominal)/3.3-3.5 GHz (peak) and 4GB RAM. The run time of DDML with conventional stacking was $2\,393$s on the full sample, while short-stacking ran in only 540s.}

\begin{table}
    \scriptsize\singlespacing\centering
    \caption{Average stacking weights with CLS}
    \label{tab:sim_advantages_weights_cls}
    \begin{tabular}{lccccccccc}
    \toprule
    \midrule
        & \multicolumn{2}{c}{Stacking}&       & \multicolumn{2}{c}{Pooled stacking} &       & \multicolumn{2}{c}{Short-stacking} \\
        \cmidrule{2-3} \cmidrule{5-6} \cmidrule{8-9}
      & $E[Y\vert X]$ & $E[D\vert X]$ &     & $E[Y\vert X]$ & $E[D\vert X]$&     & $E[Y\vert X]$ & $E[D\vert X]$ \\  
    \midrule
    \textit{Panel (A): Linear DGP and $n_b=9,915$} \\ 
    OLS&0.668&0.501&&0.738&0.573&&0.692&0.492 \tabularnewline
Lasso with CV (2nd order poly)&0.105&0.144&&0.093&0.131&&0.118&0.130 \tabularnewline
Ridge with CV (2nd order poly)&0.068&0.054&&0.050&0.040&&0.068&0.063 \tabularnewline
Lasso with CV (10th order poly)&0.027&0.073&&0.021&0.059&&0.020&0.085 \tabularnewline
Ridge with CV (10th order poly)&0.033&0.043&&0.018&0.027&&0.024&0.057 \tabularnewline
Random forest (low regularization)&0.013&0.011&&0.009&0.007&&0.009&0.008 \tabularnewline
Random forest (high regularization)&0.017&0.024&&0.012&0.018&&0.013&0.024 \tabularnewline
Gradient boosting (low regularization)&0.030&0.043&&0.024&0.034&&0.020&0.040 \tabularnewline
Gradient boosting (high regularization)&0.020&0.060&&0.020&0.070&&0.018&0.060 \tabularnewline
Neural net&0.019&0.049&&0.014&0.041&&0.018&0.043 \tabularnewline
 \\
    \textit{Panel (B): Linear DGP and $n_b=99,150$} \\  
    OLS&0.770&0.325&&0.824&0.348&&0.769&0.291 \tabularnewline
Lasso with CV (2nd order poly)&0.066&0.026&&0.055&0.012&&0.068&0.013 \tabularnewline
Ridge with CV (2nd order poly)&0.074&0.041&&0.057&0.028&&0.094&0.049 \tabularnewline
Lasso with CV (10th order poly)&0.015&0.207&&0.011&0.225&&0.011&0.198 \tabularnewline
Ridge with CV (10th order poly)&0.023&0.111&&0.018&0.095&&0.019&0.129 \tabularnewline
Random forest (low regularization)&0.003&0.003&&0.002&0.002&&0.002&0.002 \tabularnewline
Random forest (high regularization)&0.006&0.009&&0.005&0.006&&0.005&0.006 \tabularnewline
Gradient boosting (low regularization)&0.019&0.160&&0.015&0.175&&0.013&0.186 \tabularnewline
Gradient boosting (high regularization)&0.004&0.008&&0.003&0.004&&0.003&0.003 \tabularnewline
Neural net&0.019&0.110&&0.009&0.106&&0.016&0.122 \tabularnewline

          &       &       &       &       &  \\
    \textit{Panel (C): Non-linear DGP and $n_b=9,915$} \\  
    OLS&0.011&0.015&&0.007&0.012&&0.004&0.007 \tabularnewline
Lasso with CV (2nd order poly)&0.035&0.057&&0.024&0.051&&0.019&0.039 \tabularnewline
Ridge with CV (2nd order poly)&0.161&0.229&&0.192&0.268&&0.114&0.237 \tabularnewline
Lasso with CV (10th order poly)&0.053&0.080&&0.047&0.071&&0.048&0.062 \tabularnewline
Ridge with CV (10th order poly)&0.071&0.064&&0.060&0.029&&0.059&0.056 \tabularnewline
Random forest (low regularization)&0.045&0.011&&0.043&0.006&&0.043&0.005 \tabularnewline
Random forest (high regularization)&0.019&0.069&&0.010&0.065&&0.012&0.065 \tabularnewline
Gradient boosting (low regularization)&0.521&0.233&&0.548&0.251&&0.632&0.339 \tabularnewline
Gradient boosting (high regularization)&0.014&0.191&&0.005&0.203&&0.004&0.139 \tabularnewline
Neural net&0.071&0.051&&0.065&0.043&&0.064&0.049 \tabularnewline
  \\
    \textit{Panel (D): Non-linear DGP and $n_b=99,150$} \\  
    OLS&0.\phantom{000}&0.\phantom{000}&&0.\phantom{000}&0.\phantom{000}&&0.\phantom{000}&0.\phantom{000} \tabularnewline
Lasso with CV (2nd order poly)&0.\phantom{000}&0.\phantom{000}&&0.\phantom{000}&0.\phantom{000}&&0.\phantom{000}&0.\phantom{000} \tabularnewline
Ridge with CV (2nd order poly)&0.\phantom{000}&0.036&&0.\phantom{000}&0.037&&0.\phantom{000}&0.026 \tabularnewline
Lasso with CV (10th order poly)&0.\phantom{000}&0.001&&0.\phantom{000}&0.\phantom{000}&&0.\phantom{000}&0.\phantom{000} \tabularnewline
Ridge with CV (10th order poly)&0.\phantom{000}&0.036&&0.\phantom{000}&0.034&&0.\phantom{000}&0.024 \tabularnewline
Random forest (low regularization)&0.153&0.003&&0.154&0.001&&0.180&0.001 \tabularnewline
Random forest (high regularization)&0.\phantom{000}&0.060&&0.\phantom{000}&0.063&&0.\phantom{000}&0.071 \tabularnewline
Gradient boosting (low regularization)&0.845&0.853&&0.846&0.858&&0.819&0.871 \tabularnewline
Gradient boosting (high regularization)&0.\phantom{000}&0.\phantom{000}&&0.\phantom{000}&0.\phantom{000}&&0.\phantom{000}&0.\phantom{000} \tabularnewline
Neural net&0.001&0.012&&0.\phantom{000}&0.006&&0.001&0.007 \tabularnewline
 \\[-.3cm]
    \midrule
    \bottomrule
    \end{tabular}\par\medskip
    \parbox{\linewidth}{
      \scriptsize
      \textit{Notes:} The table shows the average stacking weights associated with the candidate learner for DDML with conventional stacking, pooled stacking and short-stacking. The final learner is CLS. The bootstrap sample size is denoted by $n_b$. The number of cross-fitting folds is $K=2$. Results are based on 1\,000 replications. See Table~\ref{tab:sim_advantages_bias} for more information. The final learner weights using OLS and single best are reported in Appendix Tables~\ref{tab:sim_advantages_weights_ols} and \ref{tab:sim_advantages_weights_singlebest}.
      }
\end{table}

The simulation set-up is favorable to using single-best as the final learner because there is one `true' candidate learner. However, single-best does not visibly outperform CLS, although single-best always selects the correct learner in the non-linear DGP if the sample size is sufficiently large (see Appendix Table~\ref{tab:sim_advantages_weights_singlebest}). We believe that CLS is, in practice, a good default choice. In a setting where there is a single learner that does a distinctly better job approximating the target conditional expectation function, CLS should assign a very large weight to that learner and thus approximate single-best. Otherwise, when there are several learners with similar performance or learners that perform differentially well for different observations, there are potential gains from combining the different learners.

The bias of the OLS final learner is overall similar to CLS, except when employing conventional stacking under the non-linear DGP for $n_b=9\,915$ where the average bias is almost four times as large.\footnote{Appendix Table~\ref{tab:sim_advantages_weights_ols} shows that the OLS stacking weights are often outside the unit interval. The weights associated with the neural net are particularly large (in absolute value), suggesting that OLS might be more sensitive to outliers than CLS.} The unweighted average appears sub-optimal for $n_b=99\,150$ under the non-linear DGP. Poor performance of unweighted averaging is to be expected in settings where, as in our setup, the candidate set includes learners that are not well-matched to the DGP.\footnote{By contrast, the unweighted average is known to often perform well in time-series settings \citep[e.g.,][]{clemen1989,timmermann2006}, and in particular when the optimal weights are close to equal (see \citet{Wang2023review}, Section~2.6 for a summary and wider discussion). Such a scenario is ruled out in our simulation setup, which captures the situation where the researcher does not know whether a linear or non-linear learner would be more appropriate.} We also note that the computational advantage of short-stacking with the unweighted average over short-stacking with estimated weights amounts to one (constrained) regression per conditional expectation function and is thus minimal.

The CLS weights in Table~\ref{tab:sim_advantages_weights_cls} indicate that stacking approaches successfully assign the highest weights to the estimators aligning with the data-generating process (i.e., either OLS or gradient boosting) among the ten included candidate learners, illustrating the ability to adapt to different data structures. Specifically, the stacking methods applied to the linear data-generating process assign the largest weight to linear models while they assign the largest weights to the gradient-boosting estimators and the lowest weights to estimators that impose a linear functional form on the control variables in the non-linear data-generating process.\footnote{The rates at which each candidate learner is selected by the single-best final learner are shown in Table~\ref{tab:sim_advantages_weights_singlebest} in the appendix and provide similar insights.} We conclude that DDML paired with stacking approaches reduces the burden of choice researchers face when selecting between candidate learners and specifications by allowing for the simultaneous consideration of multiple options, thus implying attractive robustness properties across a variety of data-generating processes.

\subsection{DDML and Stacking in Very Small Samples} \label{sec:simulation_very_small}

A possible concern for estimators relying on machine learning is that they might not perform well for very small samples, given that their flexibility comes at the cost of increased variance compared to parametric estimators. \citet[][henceforth WZ]{wuthrich2021} use two simulations to demonstrate that PDS lasso tends to underselect controls, which may result in a substantial small-sample bias. They also show that the bias heavily depends on the exact lasso penalty chosen (i.e., whether the plugin penalty of \citealp{Belloni2014a}, is scaled by 0.5 or 1.5), and argue in favor of OLS with appropriately chosen standard errors over PDS lasso in high-dimensional settings. 

\begin{figure}[htbp]
    \centering\scriptsize
    \caption{Replication of Figure 8 in \citet[][]{wuthrich2021}.}
    \label{fig:WZ_fig8}
    \begin{subfigure}{.48\linewidth}
    \includegraphics[width=\linewidth]{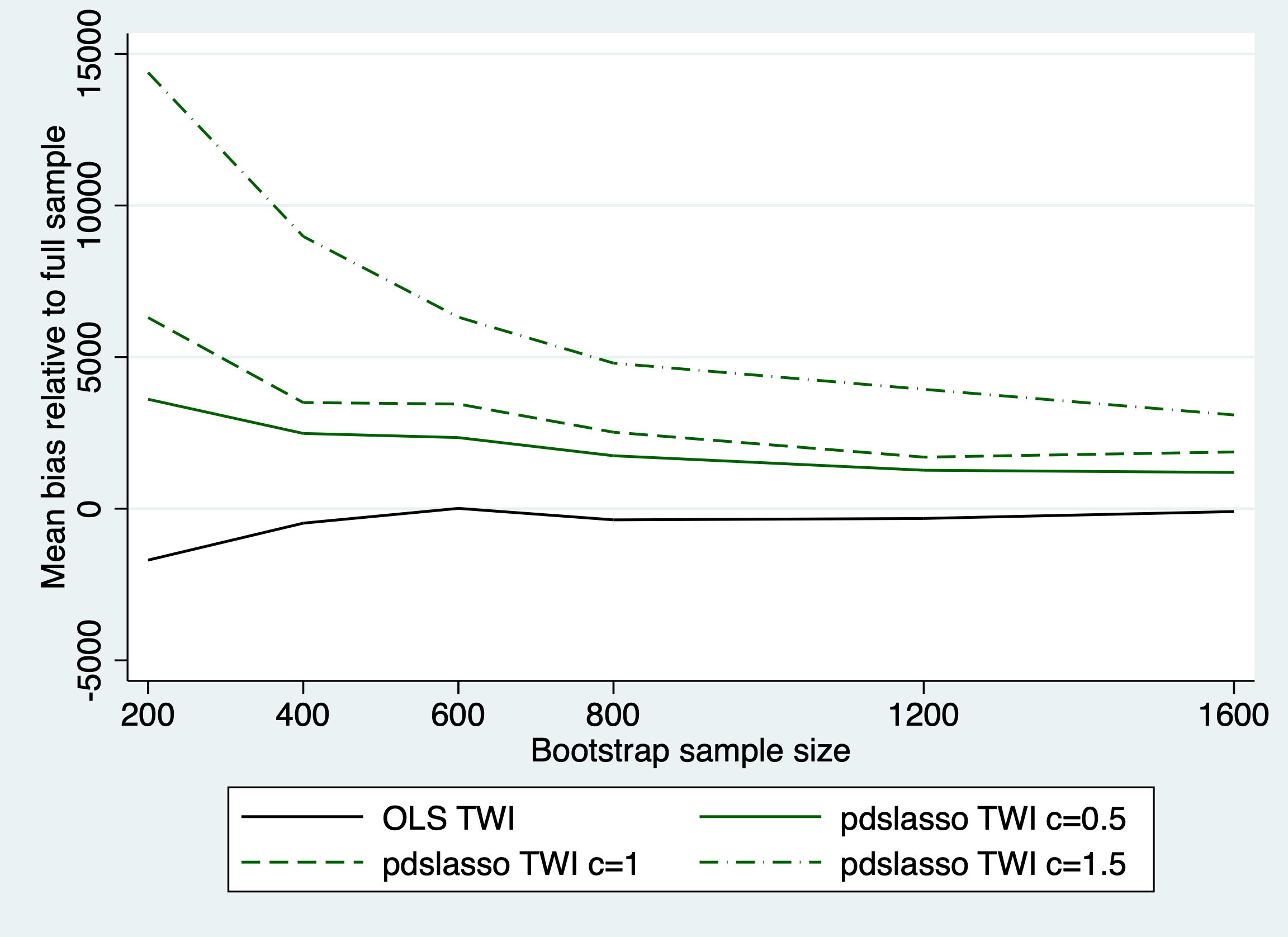}
    \caption{Bias (TWI)}
    \end{subfigure}
    \begin{subfigure}{.48\linewidth}
    \includegraphics[width=\linewidth]{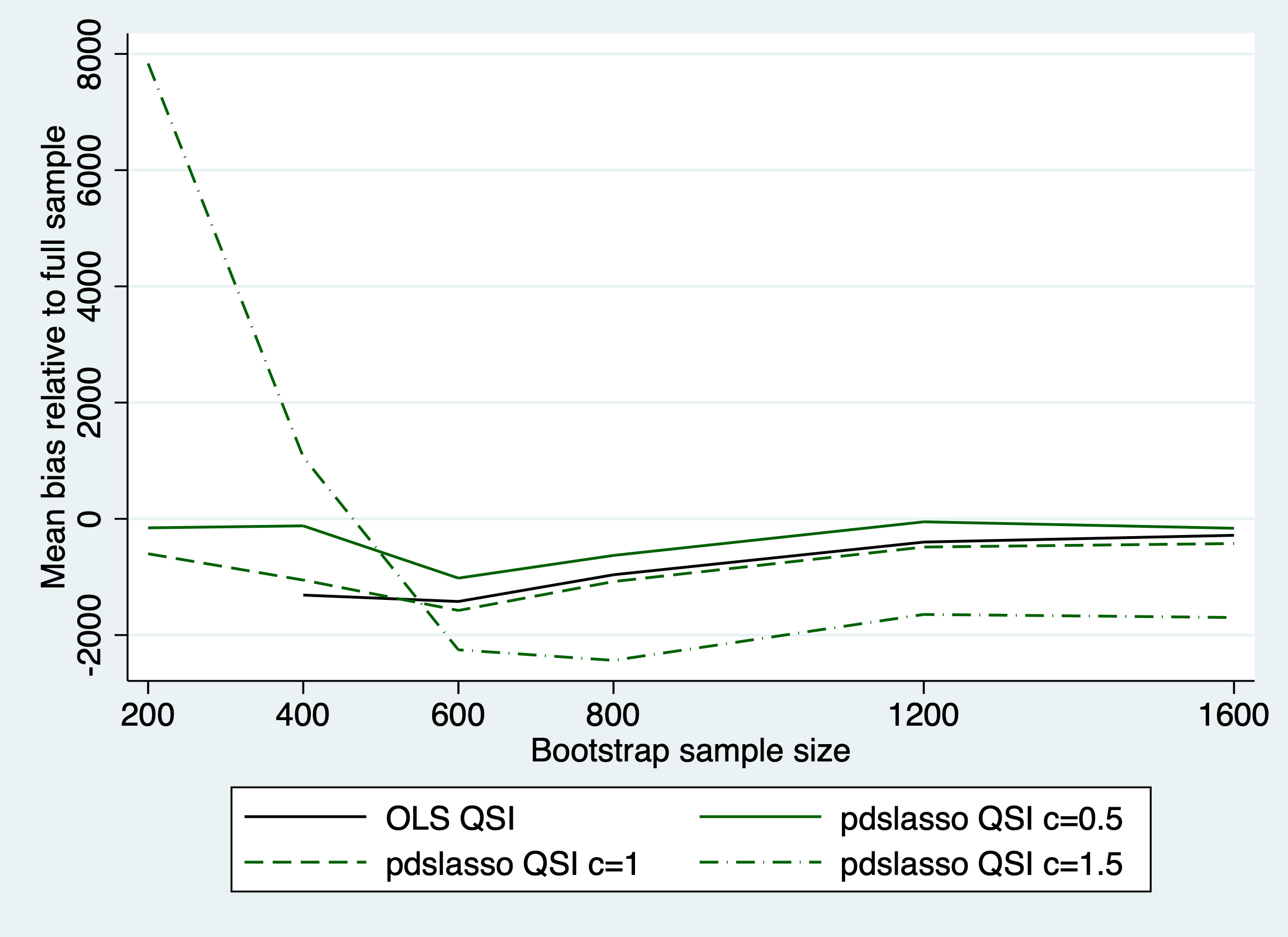}
    \caption{Bias (QSI)}
    \end{subfigure}
    \par\medskip
    \parbox{\linewidth}{\emph{Notes:} The figures report the mean bias calculated as the mean difference to the full sample estimates. Full sample estimates reported in Table~\ref{tab:WZ_fullsample}. Following WZ, we draw $1\,000$ bootstrap samples of size $n_b=\{200,400,600,800,1200,1600\}$. `TWI' indicates that the predictors have been expanded by two-way interactions. `QSI' refers to the quadratic spline \& interactions specification of \citet{Belloni2017}.
    }
\end{figure}

We revisit the 401(k) simulation set-up in WZ to assess if DDML with stacking suffers from similar issues in small samples and to compare the performance of DDML paired with stacking with PDS lasso and OLS. Following WZ, we run simulations on bootstrap samples of the data for $n_b = \{200, 400,$ $800, 1\,600\}$ and approximate the bias as the mean difference relative to the full-sample estimates ($n=9\,915$).\footnote{The full-sample estimates are reported in Table~\ref{tab:WZ_fullsample}.} WZ consider two sets of controls: two-way interactions (TWI), and quadratic splines with interactions (QSI) \citep[as in][]{Belloni2017}. The number of predictors is 167 and 272, respectively. Figure~\ref{fig:WZ_fig8} replicates the main results of WZ (Figure~8 in their paper). Panels~(a) and (b) show the bias relative to the full sample estimate for the TWI and QSI specification based on OLS and PDS lasso with tuning parameter equal to the plugin penalty of \citet{Belloni2014a} scaled by $c$ for $c \in \{0.5,1,1.5\}$. It is noteworthy that the speed at which the bootstrapped estimates converge to the full-sample estimate depends on the set of controls for the PDS lasso, but less so for OLS. While PDS lasso with $c=\{0.5,1\}$ and OLS perform similarly if QSI controls are used, PDS lasso converges much more slowly to the full-sample estimate with TWI controls. 

The DDML-stacking framework allows us to choose between, and combine, OLS and lasso with both the TWI and QSI set of controls. Another advantage of DDML over PDS lasso is that we can leverage lasso with cross-validated penalization for a fully data-driven penalization approach. Table~\ref{tab:WZ_bias} compares the performance of the full-sample estimators OLS and PDS lasso (shown in Panel A) to DDML-stacking estimators only relying on OLS and CV lasso with TWI and QSI controls as candidate learners (Panel B). We again consider conventional stacking, short-stacking and pooled stacking together with either CLS or single-best as the final learner. We set the number of cross-fitting folds to $K=10$ (but also consider $K=2$ below for comparison in Panel E).

\begin{table}[htbp]
        \scriptsize\singlespacing\centering
        \caption{Mean bias relative to full-sample estimates}
        \label{tab:WZ_bias}
                \sisetup{ 
                table-number-alignment=left,
                tight-spacing=true,
                table-format=-4.1 
                }
        \begin{tabular}{llS[table-format=5.1]S[table-format=5.1]S[table-format=5.1]S[table-format=5.1]S[table-format=5.1]S[table-format=5.1]}\toprule\toprule
        && \multicolumn{6}{c}{\it Bootstrap sample size $n_b$}\\
        \cmidrule{3-8}
        && 200 & 400 & 600 & 800 & 1200 & 1600 \\ \midrule
        \multicolumn{8}{l}{\it Panel A. Full-sample estimators}\\
        &OLS QSI&-2083.5&-910.2&-806.4&-809.9&-677.2&-626.5 \tabularnewline
&OLS TWI&-1694.5&-475.4&13.2&-366&-320.3&-91.3 \tabularnewline
&Post double Lasso QSI c=0.5&409.2&-308.9&-204&-503.1&-571.6&-354.1 \tabularnewline
&Post double Lasso QSI c=1&-179.1&-1113.5&-639.4&-1063.2&-1000.5&-523.5 \tabularnewline
&Post double Lasso QSI c=1.5&8021.3&739.9&-1526.2&-2434.4&-2255.4&-1863.5 \tabularnewline
&Post double Lasso TWI c=0.5&3611.2&2484.4&2347.2&1748.3&1270.4&1197.5 \tabularnewline
&Post double Lasso TWI c=1&6303.3&3501.1&3453.1&2523.9&1702.4&1871.8 \tabularnewline
&Post double Lasso TWI c=1.5&14386.1&8981.9&6317.9&4802.2&3939&3094.5 \tabularnewline
\\
        \multicolumn{8}{l}{\it Panel B. DDML-stacking with only OLS and CV lasso ($K=10$)}\\
        &Short-stacking: CLS&1020&-113.8&-181.1&-538.2&-575.6&-292.4 \tabularnewline
&Short-stacking: Single-best&1002.3&-122.2&-270.1&-499.7&-550.3&-197.7 \tabularnewline
&Pooled stacking: CLS&925.7&-237.3&-319.1&-628.1&-711&-370.5 \tabularnewline
&Pooled stacking: Single-best&782.3&-200.5&-358.9&-541.2&-580.2&-237.5 \tabularnewline
&Stacking: CLS&1155.8&-254.7&-266.9&-645&-633&-315.1 \tabularnewline
&Stacking: Single-best&999.5&-23.6&-184.9&-503.9&-571.1&-248.2 \tabularnewline
\\
        \multicolumn{8}{l}{\it Panel C. DDML-stacking will all candidate learners ($K=10$)}\\
        &Short-stacking: CLS&1355.1&342.2&403.3&34.2&-103.9&43.8 \tabularnewline
&Short-stacking: Single-best&669.2&113.5&144.6&-182.3&-272.6&48.9 \tabularnewline
&Pooled stacking: CLS&2849.3&1345.7&1197&383.8&-102.3&-10.6 \tabularnewline
&Pooled stacking: Single-best&724.1&-69.4&45&-250.7&-309&-19.4 \tabularnewline
&Stacking: CLS&1394.1&296.9&344.5&2.8&-168.5&56.9 \tabularnewline
&Stacking: Single-best&718.4&-47&104.3&-141.5&-318.6&42.5 \tabularnewline
\\
        \multicolumn{8}{l}{\it Panel D. DDML with candidate learners ($K=10$)}\\
        &OLS&963&-150.8&210&-161.7&-235.5&31.8 \tabularnewline
&Lasso with CV (TWI)&5948.6&3223.1&2589.1&1706.2&872.2&734.1 \tabularnewline
&Ridge with CV (TWI)&4137.3&1853.8&1617.5&951.8&657.5&879.2 \tabularnewline
&Lasso with CV (QSI)&297.5&-343.9&-311.9&-551.8&-597.1&-239.8 \tabularnewline
&Ridge with CV (QSI)&426.1&-111&85.3&-240.8&-294.4&-7.8 \tabularnewline
&Random forest (low regularization)&1852.8&618.3&709.6&259.7&7.7&95.5 \tabularnewline
&Random forest (high regularization)&9987.4&4270.1&2940.2&1919.5&1037.8&925 \tabularnewline
&Gradient boosting (low regularization)&772.3&-25&306.3&70.7&-127.2&113 \tabularnewline
&Gradient boosting (high regularization)&1060.8&94.3&564.6&292.5&44.2&228.6 \tabularnewline
&Neural net&8892.3&7481.2&6915.4&5653.2&3716.5&2224.2 \tabularnewline
\\
        \multicolumn{8}{l}{\it Panel E. DDML-stacking will all candidate learners ($K=2$)}\\
        &Short-stacking: CLS&1842.3&1078.3&-144.4&61.2&446.7&282.9 \tabularnewline
&Short-stacking: Single-best&1303.5&582.3&-436.4&-248.8&194&111.1 \tabularnewline
&Pooled stacking: CLS&2799&1471.3&159.5&209.8&572.7&508.8 \tabularnewline
&Pooled stacking: Single-best&1791.9&622.9&-542.3&-296.3&144.7&84.8 \tabularnewline
&Stacking: CLS&1924.6&1196.1&-191.2&59.4&390.3&310.9 \tabularnewline
&Stacking: Single-best&1173.4&549.6&-604.2&-285&181.8&138.3 \tabularnewline
 \\[-.3cm] \bottomrule\bottomrule
        \end{tabular}\par\medskip
        \parbox{\linewidth}{\textit{Notes:}   
        The table reports the mean bias calculated as the mean difference to the full sample estimates. Following WZ, we draw $1\,000$ bootstrap samples of size $n_b$. 
        In Panel A, we show results for the full-sample estimators OLS and PDS lasso using either two-way interactions as controls (denoted TWI) or the quadratic spline \& interactions specification of \citet[][denoted as QSI]{Belloni2017}. We scale the PDS lasso penalty by $c=0.5$, $1$ or $1.5$. 
        In Panel B, we report results for DDML with stacking approaches and only relying on OLS and CV lasso. 
        In Panel C, we consider a larger set of candidate learners. These are: OLS, CV lasso and CV ridge with either TWI or QSI controls, random forest with low regularization (8 predictors considered at each leaf split, no limit on the number of observations per node, bootstrap sample size of 70\%) or high regularization (5 splitting predictors, at least 10 observation per node, bootstrap sample size of 70\%), gradient-boosted tree with either low (500 trees, learning rate of 0.01, maximum depth of 3) or high (250 trees, learning rate of 0.01, maximum depth of 3) regularization, and a neural net with three hidden layers of size 5.
        Panel D shows results for these individual candidate learners.
        In Panels B--D, we use $K=10$ cross-fitting folds and $R=5$ cross-fitting repetitions. Panel D uses the same specifications as Panel C, but uses $K=2$. 
        }
\end{table}

Across all sample sizes, the DDML-stacking estimators strictly outperform both OLS specifications, as well as PDS lasso with TWI, and exhibit overall similar performance to PDS lasso utilizing QSI controls and $c=\{0.5,1\}$. The differences across DDML-stacking estimators are relatively minor. The CLS short-stacking weights reported in Table~\ref{tab:WZ_linear_ssw_weights}, Panel A-B, reveal that CV-lasso with QSI controls receives the largest weights, while both OLS specifications contribute jointly between nearly zero (at $n_b=200$) and only up to 15\% (for $n_b=1\,600$ and the estimation of $E[D|X]$). When selecting only a single candidate learner, CV-lasso with QSI is chosen in more than three-fourths of bootstrap iterations for the estimation of $E[Y\vert X]$ and $E[D\vert X]$ (Panel C-D in Table~\ref{tab:WZ_linear_ssw_weights}), suggesting that CV-lasso with QSI controls is strictly preferable over OLS and lasso with TWI controls in this application. This simulation exercise again highlights that relying on poorly chosen specifications that are not validated against other choices might be sub-optimal. In practice, the researcher does not know whether TWI or QSI controls perform better and whether to use OLS or lasso. Crucially, DDML paired with stacking allows for simultaneous consideration of OLS and lasso with both TWI and QSI controls and thus resolves the choice between learners and control specifications in a data-driven manner.

\begin{table}[H]
        \centering\scriptsize\singlespacing
        \caption{Short-stacking weights}
        \label{tab:WZ_linear_ssw_weights}
        \begin{tabular}{llrrrrrrr}\hline\hline
        \multicolumn{2}{l}{\it Estimator} & \multicolumn{7}{c}{\it Observations}\\
        && 200 & 400 & 600 & 800 & 1\,200 & 1\,600 & 9\,915 \\ \hline
        \multicolumn{8}{l}{\it Panel A. Constrained least squares. $E[Y|X]$, $K=10$}\\
        &OLS (TWI)&.01&.042&.062&.078&.098&.113&.013 \tabularnewline
&OLS (QSI)&0&0&.002&.008&.023&.032&.128 \tabularnewline
&Lasso with CV (TWI)&.249&.2&.196&.171&.158&.14&.214 \tabularnewline
&Lasso with CV (QSI)&.74&.758&.74&.742&.721&.716&.645 \tabularnewline
\\
        \multicolumn{8}{l}{\it Panel B. Constrained least squares. $E[D|X]$, $K=10$}\\
        &OLS (TWI)&.005&.037&.055&.074&.1&.127&.13 \tabularnewline
&OLS (QSI)&0&0&.001&.003&.011&.022&.134 \tabularnewline
&Lasso with CV (TWI)&.264&.163&.137&.119&.114&.111&.232 \tabularnewline
&Lasso with CV (QSI)&.731&.8&.807&.803&.775&.74&.504 \tabularnewline
\\
        \multicolumn{8}{l}{\it Panel C. Single-best. $E[Y|X]$, $K=10$}\\
        &OLS (TWI)&0&0&0&0&.001&0&0 \tabularnewline
&OLS (QSI)&0&0&0&0&0&0&0 \tabularnewline
&Lasso with CV (TWI)&.186&.141&.128&.112&.09&.081&0 \tabularnewline
&Lasso with CV (QSI)&.814&.859&.872&.888&.909&.919&1 \tabularnewline
\\
        \multicolumn{8}{l}{\it Panel D. Single-best. $E[D|X]$, $K=10$}\\
        &OLS (TWI)&0&0&0&0&0&0&0 \tabularnewline
&OLS (QSI)&0&0&0&0&0&0&0 \tabularnewline
&Lasso with CV (TWI)&.239&.126&.098&.079&.06&.068&.003 \tabularnewline
&Lasso with CV (QSI)&.761&.874&.902&.921&.94&.932&.997 \tabularnewline
 \\[-.3cm]\hline\hline
        \end{tabular}\par\medskip
        \parbox{\linewidth}{{\it Notes:} The table reports the stacking weights corresponding to the DDML short-stacking estimators in Figure~\ref{tab:WZ_bias}. Panel A-B use constrained least squares. Panel C-D rely on the single-best final learner. Panel A and C refer to the estimation of $E[Y|X]$; Panel C and D to the estimation of $E[D|X]$. See notes below Table~\ref{tab:WZ_bias} for more information.}
\end{table}

In the next step, we expand the set of candidate learners by two types of random forests, two types of gradient-boosted trees and a feed-forward neural net. In principle, widening the set of candidate learners increases robustness to a larger class of unknown confounding structures. We show the results in Panel~C, Table~\ref{tab:WZ_bias}. When measuring performance based on the difference to the full-sample estimates, we find there are benefits of extending the set of candidate learners for bootstrap sample sizes of $n_b=800$ or larger. The results are generally comparable across conventional, short and pooled stacking. However, single-best exhibits a lower bias for small bootstrap sample sizes vs.\ CLS, while pooled stacking with CLS appears to perform worse. The CLS weights reported in Appendix Table~\ref{tab:WZ_ssw_weights} illustrate how DDML-stacking estimators adapt to the sample size. For example, for smaller sample sizes, a larger weight is put on OLS in the estimation of $E[Y\vert X]$. In Panel~D of Table~\ref{tab:WZ_bias}, we report results for each candidate learner individually. DDML-stacking approaches perform better than most individual candidate learners and similar to the best-performing individual learner, which is DDML with CV lasso and QSI controls. In Panel E, we also show results if we reduce the number of folds to $K=2$. The performance deteriorates drastically for smaller sample sizes, indicating that---while DDML stacking appears competitive for small sample sizes---it is important to increase the number of folds to ensure larger training samples for the CEF estimators.

\begin{figure}[htbp]
        \centering\scriptsize
        \caption{Mean bias for very small sample sizes}
        \label{fig:WZ_calibrated}
        \begin{subfigure}{.48\linewidth}
        \includegraphics[width=\linewidth]{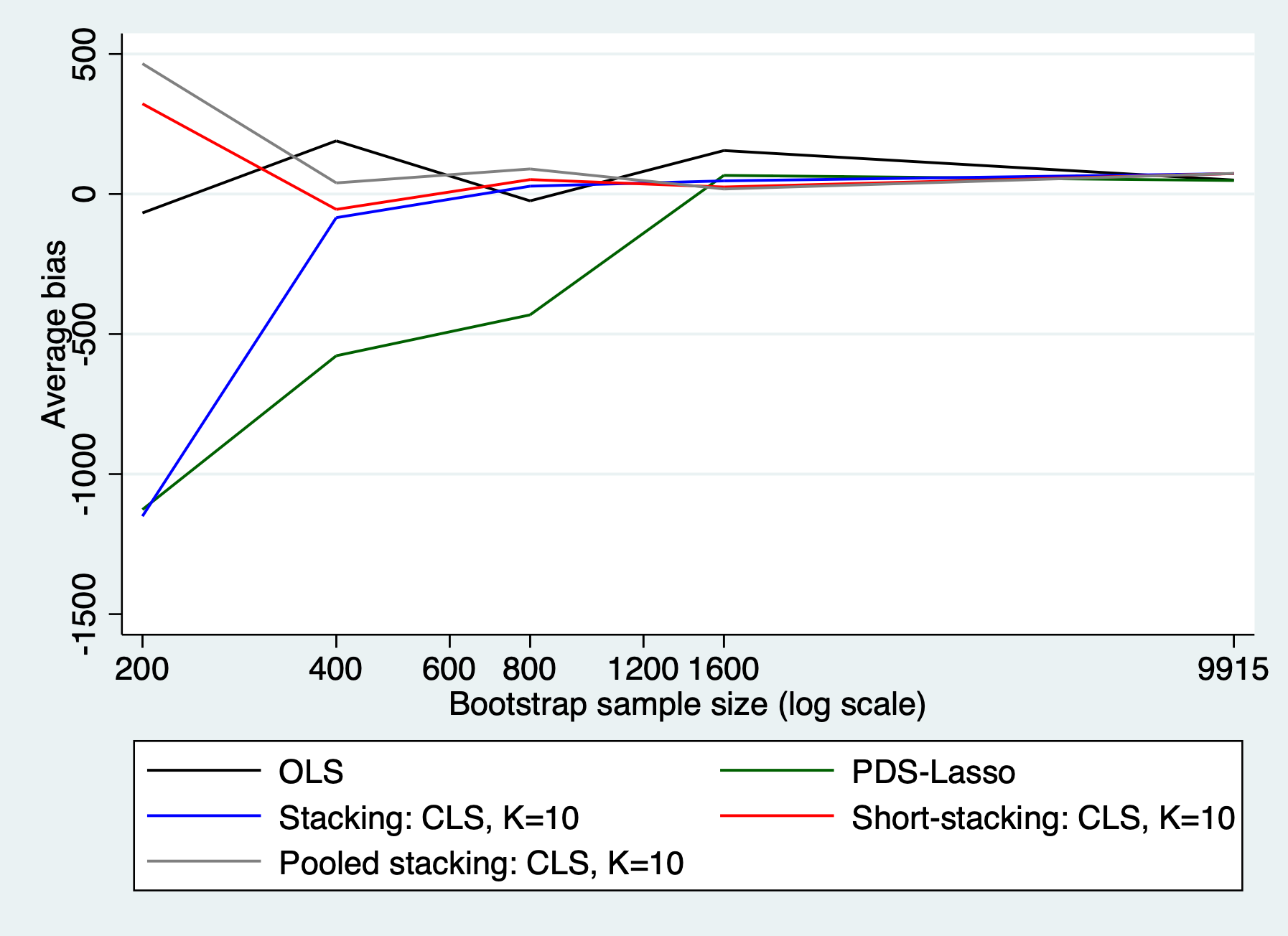}
        \caption{Linear DGP}\label{fig:WZ_calibrated_linear}
        \end{subfigure}
        \begin{subfigure}{.48\linewidth}
        \includegraphics[width=\linewidth]{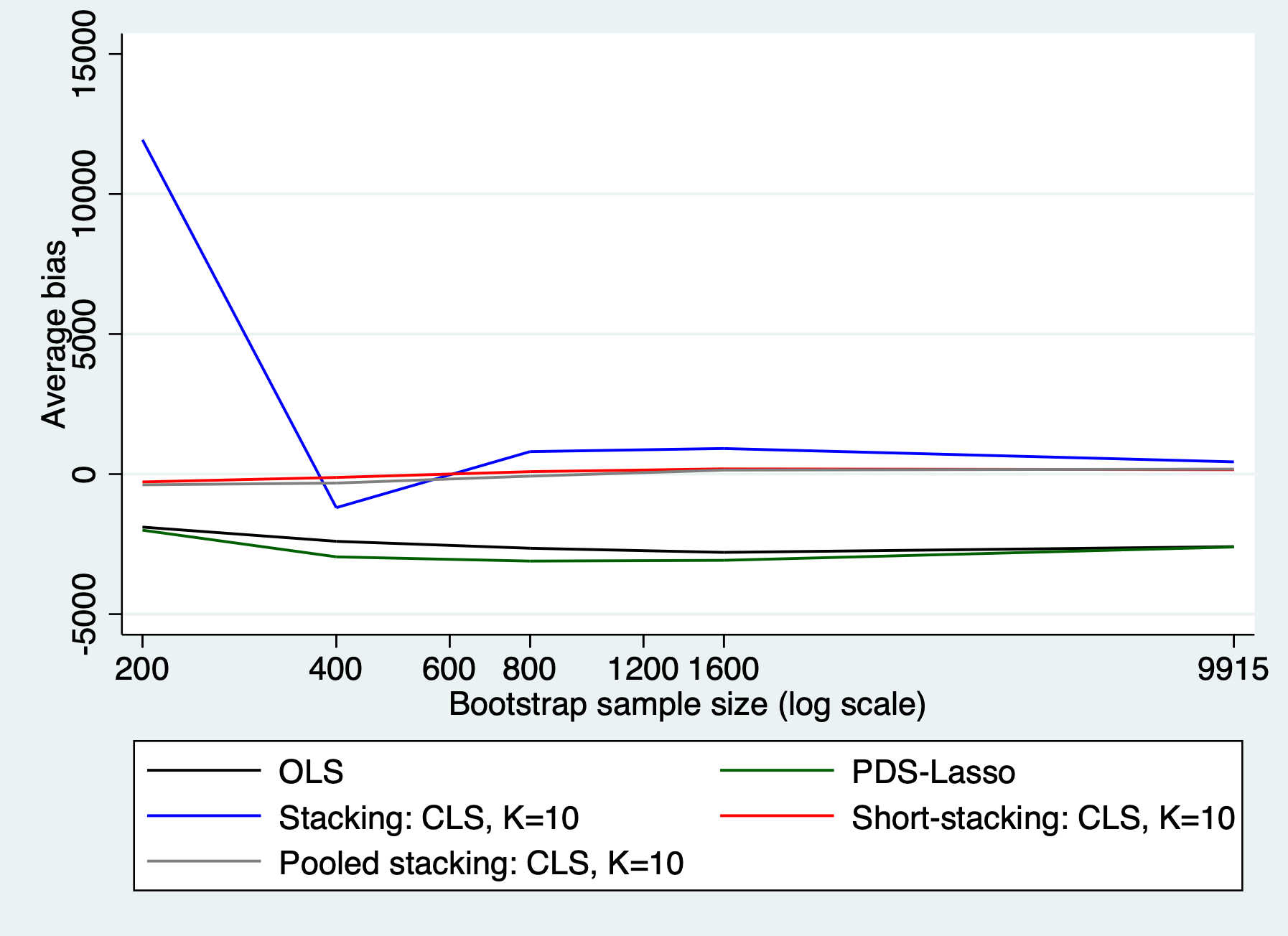}
        \caption{Non-linear DGP}\label{fig:WZ_calibrated_nonlinear}
        \end{subfigure}
        \par\medskip
        \parbox{\linewidth}{\emph{Notes:} The figure shows results from the calibrated simulation in Table~\ref{tab:sim_advantages_bias}, but with smaller bootstrap sample sizes. See table notes in Table~\ref{tab:sim_advantages_bias} for more information. Full results for bias and coverage in small samples can be found in Table~\ref{tab:sim_advantages_bias_small_linear}, \ref{tab:sim_advantages_bias_small_nonlinear} and \ref{tab:sim_advantages_coverage_small}.}
\end{figure}

A drawback of measuring the bias as the difference to the full-sample estimate is that we do not gain insights about convergence to the true parameter. We thus revisit the calibrated simulation exercise from Section~\ref{sec:advantages_stacking}, which allows us to measure the bias as the difference to the true parameter. When the DGP is linear (see Figure~\ref{fig:WZ_calibrated_linear}), DDML with short-stacking or pooled stacking and using CLS performs overall similarly to OLS. DDML with conventional stacking exhibits relatively large bias with $n_b=200$. If the true DGP is non-linear, see Figure~\ref{fig:WZ_calibrated_nonlinear}, OLS and PDS-Lasso are unable to recover the true effect, while DDML with short and pooled stacking and CLS yield reasonably close approximations of the true parameter even for small sample sizes. DDML with conventional stacking is competitive only for larger samples. We provide extensive results for mean bias and coverage rates in Tables~\ref{tab:small_sample_bias_dgp0}--\ref{tab:small_sample_cov} in the Appendix.

To conclude, the results highlight the risks of relying on inappropriate functional form assumptions. DDML paired with stacking approaches---when combined with a diverse set of candidate learners---imposes weaker conditions on the underlying data-generating process compared to relying on a single pre-selected learner. Short-stacking and pooled stacking outperform conventional stacking in small samples. We conjecture the improvement is due to short and pooled stacking imposing common weights across cross-fitting folds. 
The use of a common set of weights imposes regularization that is consistent with learner performance being stable across subsamples, which seems like a natural benchmark. For pooled stacking, this additional regularization should reduce weight variance while coming with relatively little cost in terms of bias. Because short-stacking does not make use of the additional train and test splits from conventional stacking, it has more potential to suffer from an additional over-fitting bias. When a small number of learners is considered, this additional bias should be small relative to the variance reduction obtained from not needing to estimate fold-specific stacking weights. The simulation results provide support in favor of this conjecture.

\section{Applications}\label{sec:applications}
In this section, we use two applications to illustrate how pairing DDML and stacking can increase the robustness of structural parameter estimates to the underlying structure of the data. In the first application, we estimate gaps in citations of articles in top economics journals across different gender compositions among the authors. We condition on the abstract to proxy for the content and quality of the paper and demonstrate that stacking-based DDML is a practical solution to challenging estimation problems using text data. In the second application, we revisit the UK sample of the OECD Skills Survey for Kitagawa-Oaxaca-Binder estimates of the unexplained gender wage gap where we condition on a large set of individual characteristics. Both applications pertain to the literature on gender gaps in various domains, e.g., entry to STEM programs \citep{card2021}, ICT literacy \citep{siddiq2019} or wages \citep{Strittmatter2021,bonaccolto-topfer2022}, and are methodologically also closely related to the broader literature on discriminatory attitudes towards minority groups \citep[e.g.,][]{hangartner2021monitoring}.

\subsection{Gender gap in citations}\label{sec:gender_gap_citations}

This section uses DDML with stacking to estimate a partially linear model applied to average differences in citations of articles published in top-30 economic journals from 1983 to 2020 by the gender composition of the authors. Following \citet{card2020}, we distinguish between papers with (imputed) all-male, all-female, and mixed-gender authorship.\footnote{As we explain below, we impute the gender mix of authors from the authors' names.} Instead of conditioning on hand-coded characteristics such as JEL codes, we leverage the abstract text as a proxy for the topic and quality of the article. Estimating these conditional differences is a challenging statistical problem due to the non-standard nature of text data, and researchers are faced with two key decisions when operationalizing an estimator using text data: how to encode the text data into numerical features, and how to select a suitable learner given the encoded data. Both decisions are ex-ante challenging, but also practically highly relevant as text data is becoming increasingly encountered in economic applications \citep[e.g.,][]{Gentzkow2010,chen2023gender,widmer2023media}. We show that these decisions can be consequential and that by simultaneously considering different encoding procedures and multiple learners, DDML with stacking provides a simple practical solution to both problems.

In documenting average differences in citations, the analysis presented also contributes to the broader literature on gender biases in academia \citep[e.g.,][]{lundberg2019,card2020,hengel2022}. It is well-documented that women are under-represented in academia, especially in senior positions \citep{ceci2014,lundberg2019}. A possible reason for the persistent gap in representation include is that scholarly work produced by women faces more sceptical scrutiny compared to work produced by their male counterparts \citep{hengel2022,krawczyk2016}. Higher scrutiny could be, for example, reflected at the refereeing stage when a publication decision is made and, as we examine here, after publication when scholarly work is attributed by other scholars through citations \citep{card2020,roberts2020a,grossbard2021}. 

Throughout our analysis, we focus on a descriptive characterization of the average gaps in citations across different gender compositions of the authors as given by $\theta_0$ in the partially linear model of Equation~\eqref{eq:partial_linear_model} where $Y$ denotes log-citations, $D$ is a two-dimensional vector whose first component is an indicator for all-female authorship and whose second component is an indicator for mixed-gender authorship. The vector $X$ collects the content of the abstract and a set of year-of-publication indicators. The two components of $\theta_0$ may thus be interpreted as summarizing the average relative difference in total citations between all-male and all-female authorship, and all-male and mixed-gender authorship, respectively, conditional on the article's year of publication and abstract. Throughout, we make no conditional unconfoundedness assumptions that would be necessary for causal interpretations.

We consider a sample of $27\,599$ articles that have been published between 1983--2020. The data was sourced from Scopus and is a sub-sample of the data analyzed in \citet{advani2021race}, who kindly shared their data with us. For each article, we have a record of the citation count and the authors' names, which we use to infer the authors' gender.\footnote{ We use the software \emph{Namsor}, which frequently ranks among the best-performing algorithms for gender classification using names \citep{sebo2021,krstovski2023inferring} and is widely used in academic studies \citep[e.g.,][]{Bursztyn2021,sebo2023}. In the main specification, we exclude articles of authors whose gender could not be classified with a probability of less than 70\%, but we show that results are similar when we apply thresholds of 60\% or 90\%; see Appendix Figure~\ref{fig:scopus_cites_otherthresholds}. Our sample includes 586 articles for which no citation is recorded. These were excluded from the analysis. We also provide results using the number of citations (instead of log-citations) in Appendix~Table~\ref{tab:scopus_cites_appendix}.} In the sample, 6.3\% of articles are authored by only female authors and 22.9\% have authors from both genders. 

Before turning to estimation, the text of the abstract needs to be transformed into a numerical vector. To admit estimation conditional on the content of the abstract, it is necessary to find a representation (referred to as an embedding) of the text that is lower-dimensional but captures its core meaning. An active literature in statistics and computer science provides solutions to this problem, suggesting a large variety of algorithms to construct text embeddings \citep[see the overview in][]{ash2023a}. Thus, in addition to the choice of candidate learner, researchers intent on using text data for their analysis are faced with the additional choice of embedding algorithm. To illustrate how stacking-based DDML can help support this choice, we consider two procedures for encoding the text of the abstract into numerical features: First, we consider a bag-of-word model summarizing the text as (stemmed) word counts \citep[as used in, e.g.,][]{enke2020,Esposito2023}. In our data, this results in a 211-dimensional vector of word counts for each abstract. Second, since the bag-of-word approach disregards the word order and context, we construct word embeddings generated by a pre-trained BERT model, a transformer-based large-language model \citep{devlin2018bert}. In particular, for each abstract, we extract the 768-dimensional vector of weights from the last hidden layer of the BERT model that was pre-trained on a large corpus of (uncased) English text data.\footnote{The model \texttt{bert-base-uncased} is freely available from, among others, the Python library \texttt{huggingface}.} Instead of embedding individual words, BERT attempts to reconstruct both whole sentences and the context of these sentences, making it particularly suitable to characterize the content of the abstracts. Recently, \citet{bajari2023hedonic} use BERT to construct embeddings of product descriptions on Amazon.com.

The numerical abstract embeddings are then used in several base learners. We consider OLS, PDS lasso and DDML with CV lasso, CV ridge, \texttt{XGBoost} \citep{chen2016}, random forests and a feed-forward neural net (see table notes for details).\footnote{To reduce the run time, we use regression approaches both for the estimation of $E[Y\vert X]$ and $E[D\vert X]$.} The base learners are aggregated by pairing DDML with either conventional stacking or short-stacking, and with either CLS or single-best.\footnote{We omit pooled stacking from this application since the \textsf{R} package \texttt{ddml}, which was used for this application, does not currently support pooled stacking.} The final estimator thus simultaneously combines both text embedding algorithms and machine learning algorithms.

\begin{figure}[htbp]
        \centering\singlespacing\scriptsize
        \caption{The citation gap by authors' gender composition}
        \label{fig:scopus_cites_log}
        \includegraphics[width=\linewidth]{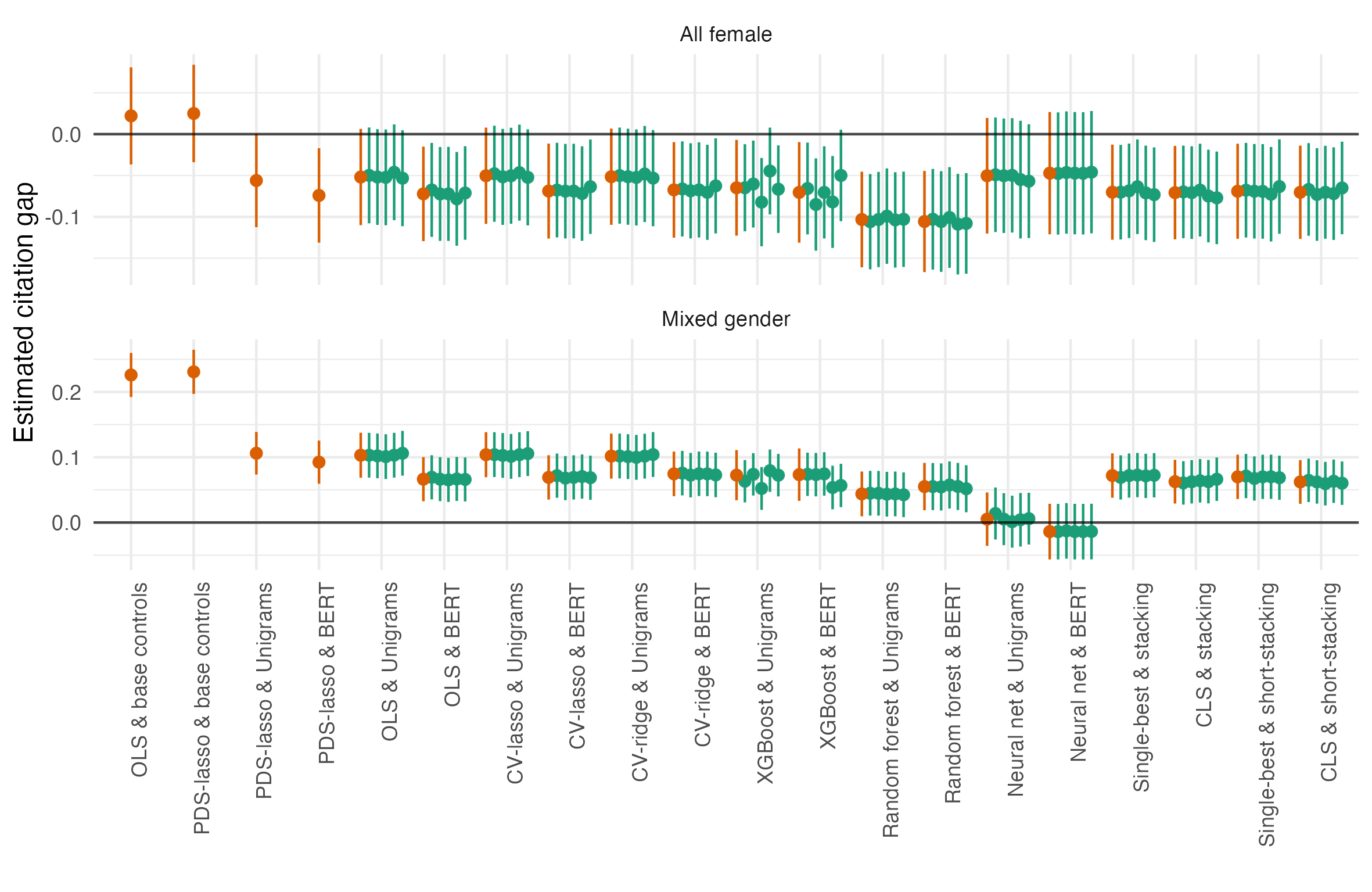}\\[-.5cm]
        \parbox{\linewidth}{\hspace*{1.4cm}$\underbracket{\hspace{2.8cm}}_{\textsf{Full-sample estimators}}$~~$\underbracket{\hspace{8.5cm}}_{\textsf{DDML with candidate learners}}$~~$\underbracket{\hspace{2.6cm}}_{\textsf{DDML and stacking}}$}\par
        \vspace*{.5cm}
        \parbox{\linewidth}{{\it Notes:}
          The figure shows estimates of $\theta_0$ summarizing average relative difference in total citations between all-male and all-female authorship, and all-male and mixed-gender authorship, respectively, conditional on the article's year of publication and abstract. Error-bars show heteroskedasticity-robust 95\% confidence intervals. The sample mean (and standard deviation) of citation counts, all female and mixed gender are 99.8 (254.0), 0.062 (0.242) and 0.229 (0.420), respectively. We consider the following estimators: OLS, PDS lasso and DDML with the following candidate learners: OLS, CV ridge, CV lasso, XGBoost (using 500 trees, learning rate of 0.3, maximum depth of 6), random forest (using 500 trees) and feed-forward neural net (early stopping with 15 rounds, 0.5 dropout, 0.1 learning rate, 0.1 validation split, 50 epochs, 500 batch size and 3 hidden layers of size 10). Finally, we pair DDML with either conventional stacking or short-stacking, and with either CLS or single-best as the final learner based on the above candidate learners. Throughout, we use five cross-fitting repetitions, five cross-validation folds and five cross-fitting folds. Results from each cross-fitting replication are illustrated in green, and median aggregates across the cross-fitting replications are shown in orange. The sample includes $29\,185$ articles published between 1983--2020 in top-30 economics journals. A tabular version is provided in Table~\ref{tab:scopus_cites_appendix}. Authors that could not be assigned a gender with less than 70\% probability are excluded from the analysis, but we show results based on thresholds of 60\% and 90\% in Appendix Figure~\ref{fig:scopus_cites_otherthresholds}.}
\end{figure}

Figure~\ref{fig:scopus_cites_log} shows estimates of the average relative difference in total citations between all-male and all-female authorship (top-panel) and all-male and mixed-gender authorship (bottom-panel), respectively, for different control specifications and estimators.  When we only condition on the publication year, the citation penalty for all-female authorship is close to zero, while there is a large positive effect of $+22.6$\% ($s.e.=1.7$) for mixed-gender authorship. We next employ PDS lasso to add the abstract text either in the form of word counts or as BERT features. Using the latter, the citation gap increases to $-7.4$\% ($2.9$) for articles with all-female authorship, while the average relative difference of articles with mixed-gender authorship reduces to $+9.2$\% ($1.7$). The estimates are qualitatively similar when using word counts instead of BERT features. 

In the figure, we also show five cross-fitting repetitions of pairing DDML with each candidate learner (in green) and the median aggregates over these repetitions (in orange). There are considerable differences across DDML estimators, with the median estimates of the citation gap ranging between $-4.7$\% ($3.8$) and $-10.6$ ($3.1$) for articles with all-female authorship and between $-1.4$ (2.2) and $+10.4$ ($1.8$) for articles with mixed-gender authorship, highlighting that different candidate learner specifications can yield vastly different effect sizes. These stark differences emphasize the need to choose and tune CEF estimators carefully. Without thoroughly validating each candidate learner, judging which results are more credible is difficult. Furthermore, it is noteworthy that some candidate learners, specifically those based on XGBoost in this example, exhibit substantial instability across cross-fitting repetitions.

We show results from pairing DDML and stacking approaches on the right-hand side of the same figure. Relative to the DDML estimates based on the individual candidate learners, the stacking approaches yield lower variability over cross-fitting repetitions. All four stacking-based approaches agree on an average relative difference in citations of around $-7.0$\% ($2.9$) for articles with all-female authorship and suggest a citation advantage of between $+6.2$ ($1.7$) and $+7.0$ ($1.7$) for articles with mixed-gender authorship.

\begin{table}[htb]
        \centering\scriptsize\singlespacing
        \caption{Stacking weights in the gender citation gap application.}
        \label{tab:scopus_cites_log_weights_mspe}
        \begin{tabular}{llcccccccccccc}\toprule\toprule
        && &\multicolumn{2}{c}{\it Citations} & &\multicolumn{2}{c}{\it All female} & & \multicolumn{2}{c}{\it Mixed gender}  \\
        && & \it Conv. & \it Short & &  \it Conv. & \it Short & &\it Conv. &\it Short  \\
        \midrule
        \multicolumn{8}{l}{\it Panel A. Stacking and short-stacking weights} \\
        \partialinput{4}{15}{scopus_cites/weights_log.tex} \\
        && &\multicolumn{2}{c}{\it Citations} & &\multicolumn{2}{c}{\it All female} & & \multicolumn{2}{c}{\it Mixed gender}  \\
        \midrule
        \multicolumn{8}{l}{\it Panel B. Mean-squared prediction error}\\
       \partialinput{4}{15}{scopus_cites/mspe_log_joined.tex}  \\[-.3cm]
        \bottomrule\bottomrule
        \end{tabular}\par\medskip
        \parbox{\linewidth}{\emph{Notes:} Panel A shows stacking weights for conventional stacking (labelled `Conv.') and short-stacking (labelled `Short') by candidate learners and by variable. Panel~B reports the mean-squared prediction error. The final learner is constrained least squares. The stacking weights and mean-squared prediction errors are averaged over cross-fitting repetitions. Treatment variables are an indicator for all-female authors and mixed-gender authors.}
\end{table}

Table~\ref{tab:scopus_cites_log_weights_mspe} shows the stacking weights of conventional and short-stacking with constrained least squares as the final learner along with mean-squared prediction errors. The stacking estimators assign small weights to learners exhibiting a relatively large MSPE and large variability over cross-fitting repetitions. For example, in the CEF estimation of log citations, the neural nets have an MSPE that is around 50\% larger than that of other learners. Stacking assigns, as desired, zero weights to the neural nets, whereas OLS leveraging BERT as one of the best-performing learners receives the largest weights. It is noteworthy that the stacking weights often vary markedly across CEFs, highlighting that there is no reason to assume that the same learner is best suited for estimating both $E[Y\vert X]$ and $E[D\vert X]$. This insight is especially important since most estimation approaches (including OLS and PDS lasso) impose the same structure for each CEF. 

The results on the citation gaps in top economic journals conditional on the content of the abstract are consistent with a citation penalty for all-female authored articles, possibly due to a higher degree of skepticism towards all-female author teams compared to all-male author teams. However, similar to \citet{card2020} and \citet{maddi2021}, the estimates also suggest a conditional citation advantage of articles with mixed-gender authorship.

\subsection{Gender gap in wages}\label{sec:gwg}
The gap in wages between men and women is a central measure of economic gender equality and has been the focus of an extensive empirical literature \citep[see, e.g., the review in][]{blau2017}. The classic approach to estimating the unexplained gender wage gap relies on a linear version of the Kitagawa-Oaxaca-Binder decomposition (\citealp{kitagawa1955,oaxaca1973,blinder1973}; for an overview, see \citealp{fortin2011}). Several recent articles by \citet{bonaccolto-topfer2022}, \citet{Strittmatter2021}, \citet{boheim2021decomposition} and \citet{bach2023}, among others, focus instead on semi-parametric decompositions of the wage gap leveraging more flexible machine learning algorithms. Much of this literature focuses, however, on lasso-based approaches, even though there is no apparent reason to favor sparsity-based approaches over learners relying on other regularization assumptions. In contrast to the recent literature that primarily focuses on lasso-based approaches, we consider a diverse set of candidate learners and aggregate them via stacking. 

The parameter of interest in this application is the unexplained gender wage gap, which is the expected difference in wages after conditioning on observed characteristics. Formally, \begin{align*}
    \theta_0 \equiv E\left[E\left[Y \vert D = 1, X\right] - E\left[Y \vert D = 0, X\right] \vert D = 1\right],
\end{align*}
where $Y$ denotes the logarithm of wages, $D$ is an indicator equal to one for women, and $X$ is a vector of potentially many individual characteristics. The parameter is well-defined if $P(D=1\vert X ) > 0$ with probability 1.\footnote{As in the previous section, we focus our analysis on a descriptive parameter of interest and do not make conditional unconfoundedness assumptions that would be necessary for causal interpretations.}

In the absence of functional form assumptions, estimation of $\theta_0$ is a challenging statistical problem due to its dependence on unknown conditional expectation functions that need to be nonparametrically estimated. Analogous to the DDML estimator for the partially linear model outlined in Section \ref{sec:theory}, we consider estimation of $\theta_0$ via the split-sample analogue of the efficient score function for $\theta_0$\footnote{See, e.g., equation (5.4) in \citet{Chernozhukov2018}.} -- i.e., \begin{align*}
      \hat{\theta}_n &=\frac{1}{n}\sum_{i=1}^n\left(\frac{D_i(Y_i-\hat{g}_{I^c_{k_i}}(0,\bm{X}_i))}{\hat{p}_{I^c_{k_i}}} - \frac{\hat{m}_{I^c_{k_i}}(\bm{X}_i)(1-D_i)(Y_i-\hat{g}_{I^c_{k_i}}(0,\bm{X}_i))}{\hat{p}_{I^c_{k_i}}(1-\hat{m}_{I^c_{k_i}}(\bm{X}_i))}\right) ,
\end{align*}
where $\hat{g}_{I^c_{k}}$ and $\hat{m}_{I^c_{k}}$ are cross-fitted estimators for $g_0(D, X)\equiv E[Y\vert D, X]$ and $m_0(X) \equiv E[D\vert X]$, and {$\hat{p}_{I^c_{k}}$ is a cross-fitted estimator of $P(D=1)$. 

Following \citet{forshaw2024gender}, we take the data for this application from the UK sample of the OECD Skills Survey, which was collected in 2011-12 and comes with a rich set of covariates, including age, experience, education, occupation, and industry. The final data includes $4\,836$ British respondents. We specify three sets of control variables. The \emph{simple} set of controls only includes a selection of essential covariates: age (in levels and squared), years of education, a literacy and numeracy test score, years of tenure in the current job (in levels and squared), education level, hours worked per week, and number of children. The \emph{base} set of controls adds, among others, management level, age of children, and parents' education level.\footnote{ The base set adds the following variables to the reduced set: area of study, part of larger organization, management position, type of contract, job satisfaction, health status, living with a partner, age of youngest child, immigration age, mother's and father's highest level of education, immigration status of parents, informal job-related education in last 12 months, informal non-job-related education in last 12 months.} Furthermore, we interact age and tenure with all categorical covariates. The \emph{extended} set of controls comprises all variables and interacts each continuous covariate with each categorical covariate. 

We include a diverse set of candidate learners to allow for a high level of flexibility. We employ regression approaches for the estimation of the CEF of log wages (i.e., $E[Y\vert D, X]$) and classification approaches for the CEF estimation of gender (i.e., $E[D\vert X]$). Our candidate learners are linear (or logistic) regression with the simple and base set of controls; linear (or logistic) CV-lasso and CV-ridge with the base and extended set of controls; three random forests with 500 regression (or classification) trees and minimum leaf sizes of 1, 50 and 100; two types of gradient-boosted regression (or classification) trees with and without early stopping (500 trees, maximum depth of 3); two feed-forward neural nets with hidden layer sizes of $(40,20,1,20,50)$ and $(30,30,30)$ and early stopping.  Finally, we aggregate the candidate learners via conventional, short and pooled stacking, using either CLS or single-best as the final learner.

\begin{figure}[htbp] \scriptsize\singlespacing
        \caption{Unexplained gender wage gap}\label{fig:gwg}
        \centering
        \vspace*{-.3cm}
        \includegraphics[width=.9\linewidth]{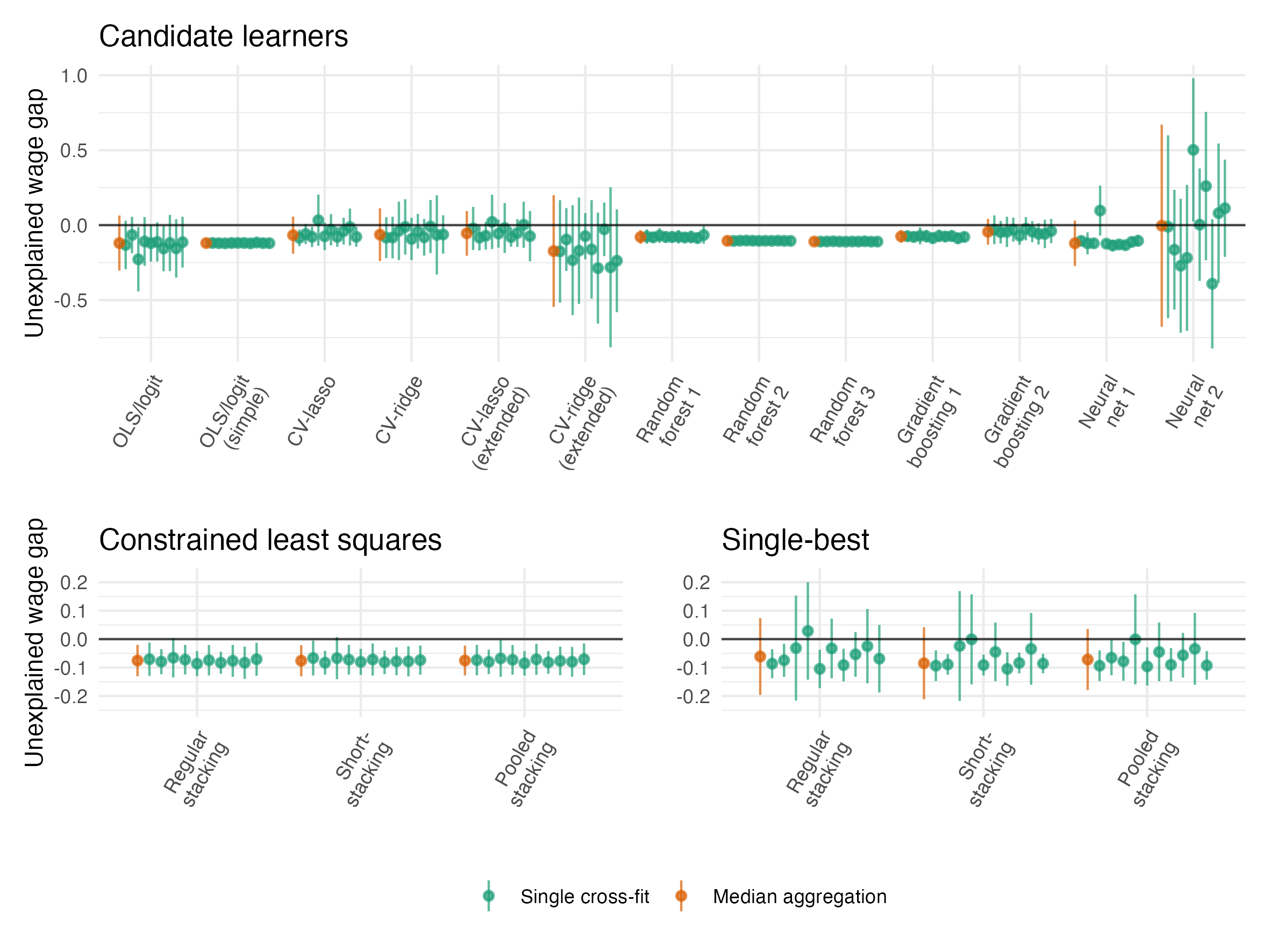}\par\medskip
        \parbox{\linewidth}{{\it Notes:}  The figure reports DDML estimates of the unexplained gender wage gap based on several different learners. 95\% heteroskedasticity-robust confidence intervals are shown. The sample mean (and standard deviation) of log earnings and gender are 2.768 (0.579) and 0.583 (0.493), respectively. The candidate learners are OLS (for the outcome equation) and logit (for the propensity scores) with the base and simple set of controls; CV-lasso and CV-ridge with the base and extended set of controls; three random forests with 500 trees and minimum leaf sizes of 1, 50 and 100; two gradient-boosted trees with and without early stopping (500 trees, with and without early stopping after 10 iterations, maximum depth of 3); two feed-forward neural nets with hidden layer sizes of $(40,20,1,20,50)$ and $(30,30,30)$ and early stopping. We report results for the individual base learners in the top panel. In the bottom panels, we show DDML paired with conventional, short and pooled stacking based on the base learners and with either CLS (left panel) or single-best (right panel) as the final learner. We use 10 cross-fitting folds and 10 cross-fitting repetitions. Results from each cross-fitting replication are illustrated in green, and median aggregates across the cross-fitting replications are shown in orange. A tabular version is provided in Table~\ref{tab:gwg_results_1}-\ref{tab:gwg_results_2}.}
\end{figure}

Figure~\ref{fig:gwg} reports results for individual candidate learners (on the top) and stacking approaches (on the bottom). We show results from 10 cross-fitting repetitions (in green) and the median aggregates (in orange). We again find that some candidate learners exhibit substantial variability over cross-fitting repetitions, which is also reflected in the large median-aggregate standard errors. The variability is especially large for CV-ridge with the extended set of controls and the neural nets, which are the candidate learners exhibiting the largest MSE (see Appendix Table~\ref{tab:gwg_weights}). The stacking results are, in contrast, relatively stable over cross-fit repetitions when using CLS as the final learner. Interestingly, the results when relying on single-best as the final learner are noticeably more variable than when using CLS, indicating that a combination of candidate learners seems to better fit the data than a single learner. The stacking weights and MSE in Appendix Table~\ref{tab:gwg_weights} confirm that there is no single candidate learner dominating the others. The instability of the single-best final learner is reflected in the stacking standard errors. Given this potential for instability of choosing a single candidate learner, we recommend favoring constrained least squares over single-best if one is not confident that one of the chosen learners will be significantly better than the rest, and thus stably selected, which seems likely to be the most common setting in practice.

\section{Conclusion}\label{sec:conclusion}
This article assesses the performance of DDML estimators in realistic settings using applications and simulation studies calibrated to real economic data. We highlight that estimators of structural parameters based on single pre-selected (machine) learners can be highly sensitive to the underlying structure of the data and/or poor tuning, and we show that pairing DDML with stacking can help alleviate these concerns, provided that a sufficiently diverse set of candidate learners is considered. 

We discuss pairing DDML with conventional stacking but also suggest two alternative stacking approaches: Short-stacking, which substantially reduces the computational burden by leveraging the cross-fitting naturally arising in the computation of DDML estimates, and pooled stacking, which decreases the variance of the stacking estimator by imposing common stacking weights over cross-fitting folds. In our simulations, both strategies are competitive with conventional stacking in settings with large and moderate sample sizes and are better in small samples. The advantages of short-stacking are particularly worth highlighting, given its substantially lower computational cost.

A key advantage of the DDML-stacking approach is that it accommodates both traditional parametric and nonparametric specifications by allowing simultaneous consideration of, for example, OLS with several sets of controls, sparsity-based learners, tree-based ensembles and neural networks. In this sense, researchers are not forcibly deviating from standard (often linear) specifications unless the data suggests there is reason to. While machine-learning-based causal methods may yield fundamentally different results from linear regression only in specific examples, the additional robustness to unexpected structures in the data thus seems to come at relatively little cost.

\printbibliography

\newpage\singlespacing

\pagestyle{empty}


\appendix
\pagestyle{plain}
\setcounter{page}{1}
\setcounter{figure}{0}
\setcounter{table}{0}

\renewcommand\thefigure{\thesection.\arabic{figure}} 
\renewcommand\thetable{\thesection.\arabic{table}} 

\begin{center}
	\Large\bfseries
	Supplementary material
\end{center}

\newcommand{\bracketr}{]}
\newcommand{\bracketl}{[}

\section{The benefits of pairing DDML and stacking}

\begin{table}[htbp]
    \scriptsize
    \centering
    \caption{Mean-squared prediction error}
    \label{tab:sim_advantages_mspe}
    \begin{tabular}{llccccc}
    \toprule
    \midrule
    & & \multicolumn{2}{c}{$n_b = 9\,915$} &       & \multicolumn{2}{c}{$n_b = 99\,150$} \\
    \cmidrule{3-4}\cmidrule{6-7}\multicolumn{2}{l}{\textit{Panel (A): Linear DGP}} & $E[Y\vert X]$ & $E[D\vert X]$ &       & $E[Y\vert X]$ & $E[D\vert X]$ \\
    \midrule
    \multicolumn{2}{l}{\it ~~Candidate learners}\\
    &OLS&3.093&0.200&&3.088&0.200 \tabularnewline
&Lasso with CV (2nd order poly)&3.095&0.200&&3.088&0.200 \tabularnewline
&Ridge with CV (2nd order poly)&3.100&0.200&&3.089&0.200 \tabularnewline
&Lasso with CV (10th order poly)&3.298&0.202&&3.095&0.200 \tabularnewline
&Ridge with CV (10th order poly)&3.423&0.206&&3.093&0.200 \tabularnewline
&Random forest (low regularization)&3.613&0.233&&3.698&0.239 \tabularnewline
&Random forest (high regularization)&3.183&0.205&&3.197&0.207 \tabularnewline
&Gradient boosting (low regularization)&3.131&0.201&&3.102&0.200 \tabularnewline
&Gradient boosting (high regularization)&3.151&0.201&&3.137&0.201 \tabularnewline
&Neural net&3.227&0.204&&3.153&0.200 \tabularnewline

          &       &       &       &       &  \\
    \multicolumn{2}{l}{\textit{Panel (B): Non-Linear DGP}} & $E[Y\vert X]$ & $E[D\vert X]$ &       & $E[Y\vert X]$ & $E[D\vert X]$ \\
    \midrule
    \multicolumn{2}{l}{\it ~~Candidate learners}\\
    &OLS&3.681&0.203&&3.672&0.203 \tabularnewline
&Lasso with CV (2nd order poly)&3.480&0.201&&3.449&0.200 \tabularnewline
&Ridge with CV (2nd order poly)&3.479&0.201&&3.449&0.200 \tabularnewline
&Lasso with CV (10th order poly)&6.161&0.223&&3.422&0.200 \tabularnewline
&Ridge with CV (10th order poly)&7.431&0.230&&3.424&0.200 \tabularnewline
&Random forest (low regularization)&3.789&0.231&&3.515&0.236 \tabularnewline
&Random forest (high regularization)&3.588&0.204&&3.251&0.205 \tabularnewline
&Gradient boosting (low regularization)&3.345&0.200&&3.095&0.198 \tabularnewline
&Gradient boosting (high regularization)&3.399&0.200&&3.216&0.199 \tabularnewline
&Neural net&3.694&0.205&&3.510&0.200 \tabularnewline
  \\[-.3cm]
    \midrule
    \bottomrule
    \end{tabular}\par\medskip
        \parbox{.74\linewidth}{{\it Notes:} The table shows the mean-squared prediction error of each candidate learner from the simulation example in Section \ref{sec:advantages_stacking}. The bootstrap sample size is $n_b=9\,915$ or $99\,150$. Results are based on $1\,000$ replications. See Table~\ref{tab:sim_advantages_bias} for more information.
    }
\end{table}

\begin{table}[htbp]
        \centering\scriptsize\singlespacing
        \caption{Bias in the Linear and Non-Linear DGP}
        \sisetup{ 
            table-number-alignment=left,
            tight-spacing=true,
            table-format=-4.1 
        }
        \label{tab:sim_advantages_bias_se}
            \begin{tabular}{rlrrcrrcrrcrr}
        \toprule
        \midrule
        && \multicolumn{5}{c}{\textit{Panel (A): Linear DGP}} & &\multicolumn{5}{c}{\textit{Panel (B): Non-linear DGP}}\\
        \cmidrule{3-7}\cmidrule{9-13}
              &       & \multicolumn{2}{c}{$n_b = 9\,915$} &       & \multicolumn{2}{c}{$99\,150$}
              &       & \multicolumn{2}{c}{$n_b = 9\,915$} &       & \multicolumn{2}{c}{$99\,150$}\\
        \cmidrule{3-4}\cmidrule{6-7}\cmidrule{9-10}\cmidrule{12-13}
        && \multicolumn{1}{c}{Bias}  & \multicolumn{1}{c}{s.e.}      &       & \multicolumn{1}{c}{Bias}  & \multicolumn{1}{c}{s.e.}   & & \multicolumn{1}{c}{Bias}  & \multicolumn{1}{c}{s.e.}       &       & \multicolumn{1}{c}{Bias}  & \multicolumn{1}{c}{s.e.}   \\ \midrule
              \multicolumn{4}{l}{Full sample:}\\
              \partialinput{1}{2}{Simul/sim_Advantages/sim_output_folds2_se}
               \multicolumn{4}{l}{DDML methods:}\\
               \multicolumn{4}{l}{\it ~~Candidate learners}\\
              \partialinput{3}{12}{Simul/sim_Advantages/sim_output_folds2_se}
              \multicolumn{4}{l}{\it ~~Stacking approaches}\\
              \partialinput{13}{24}{Simul/sim_Advantages/sim_output_folds2_se}
        \midrule
        \bottomrule
        \end{tabular}%
        \par\medskip
      \parbox{\linewidth}{%
      \textit{Notes:} The table reports mean bias and associated standard errors (s.e.) for the listed estimators from the simulation example in Section \ref{sec:advantages_stacking}.
      Results are based on $1\,000$ replications. See Table~\ref{tab:sim_advantages_bias} for more information.
      }
\end{table}

\begin{table}[htbp]
    \scriptsize\singlespacing\centering
    \caption{Average stacking weights using OLS as the final learner}
    \label{tab:sim_advantages_weights_ols}
    \begin{tabular}{lrrrrrrrrr}
    \toprule
    \midrule
        & \multicolumn{2}{c}{Stacking}&       & \multicolumn{2}{c}{Pooled stacking} &       & \multicolumn{2}{c}{Short-stacking} \\
        \cmidrule{2-3} \cmidrule{5-6} \cmidrule{8-9}
         & $E[Y\vert X]$ & $E[D\vert X]$ &     & $E[Y\vert X]$ & $E[D\vert X]$&     & $E[Y\vert X]$ & $E[D\vert X]$ \\
         \midrule
     \textit{Panel (A): Linear DGP and $n_b=9,915$}    \\
    OLS&1.161&0.932&&1.014&0.815&&0.923&0.686 \tabularnewline
Lasso with CV (2nd order poly)&0.011&0.034&&0.066&0.085&&0.154&0.079 \tabularnewline
Ridge with CV (2nd order poly)&--0.157&--0.187&&--0.091&--0.137&&--0.080&--0.080 \tabularnewline
Lasso with CV (10th order poly)&--0.046&0.076&&--0.027&0.069&&--0.041&0.122 \tabularnewline
Ridge with CV (10th order poly)&0.\phantom{000}&--0.006&&0.\phantom{000}&0.003&&0.017&0.022 \tabularnewline
Random forest (low regularization)&0.001&--0.005&&0.001&--0.006&&0.003&--0.006 \tabularnewline
Random forest (high regularization)&--0.014&--0.006&&--0.010&--0.002&&--0.011&0.008 \tabularnewline
Gradient boosting (low regularization)&--0.060&--0.108&&--0.045&--0.096&&--0.039&--0.094 \tabularnewline
Gradient boosting (high regularization)&0.096&0.260&&0.090&0.259&&0.076&0.257 \tabularnewline
Neural net&--195.245&0.009&&0.008&0.034&&0.004&0.033 \tabularnewline
 \\ \\
    \textit{Panel (B): Linear DGP and $n_b=99,150$} \\ 
    OLS&1.292&0.935&&1.130&0.769&&1.008&0.677 \tabularnewline
Lasso with CV (2nd order poly)&--0.103&--0.497&&--0.036&--0.378&&0.033&--0.398 \tabularnewline
Ridge with CV (2nd order poly)&--0.170&--0.053&&--0.093&--0.019&&--0.044&0.056 \tabularnewline
Lasso with CV (10th order poly)&0.065&0.407&&0.023&0.368&&0.009&0.335 \tabularnewline
Ridge with CV (10th order poly)&--0.071&--0.118&&--0.022&--0.068&&--0.003&--0.034 \tabularnewline
Random forest (low regularization)&--0.000&--0.001&&--0.000&--0.001&&0.\phantom{000}&--0.001 \tabularnewline
Random forest (high regularization)&--0.001&--0.002&&--0.000&--0.001&&--0.000&--0.001 \tabularnewline
Gradient boosting (low regularization)&--0.034&0.158&&--0.021&0.177&&--0.020&0.224 \tabularnewline
Gradient boosting (high regularization)&0.034&0.072&&0.027&0.058&&0.022&0.020 \tabularnewline
Neural net&--12.267&0.110&&--0.005&0.108&&--0.003&0.133 \tabularnewline

          &       &       &       &       &  \\
      \textit{Panel (C): Non-Linear DGP and $n_b=9,915$}   \\
    OLS&0.013&0.050&&0.004&0.038&&--0.048&0.022 \tabularnewline
Lasso with CV (2nd order poly)&--0.118&--0.319&&--0.166&--0.212&&0.033&--0.347 \tabularnewline
Ridge with CV (2nd order poly)&0.365&0.528&&0.427&0.457&&0.146&0.548 \tabularnewline
Lasso with CV (10th order poly)&--0.014&0.126&&0.015&0.118&&0.091&0.088 \tabularnewline
Ridge with CV (10th order poly)&0.060&0.028&&0.040&--0.003&&--0.011&0.055 \tabularnewline
Random forest (low regularization)&0.052&--0.016&&0.056&--0.017&&0.057&--0.016 \tabularnewline
Random forest (high regularization)&--0.098&0.059&&--0.099&0.065&&--0.110&0.064 \tabularnewline
Gradient boosting (low regularization)&1.096&0.029&&1.133&0.059&&1.358&0.159 \tabularnewline
Gradient boosting (high regularization)&--0.531&0.526&&--0.586&0.508&&--0.716&0.426 \tabularnewline
Neural net&--72.340&0.027&&0.121&0.046&&0.120&0.054 \tabularnewline
 \\ 
    \textit{Panel (D): Non-Linear DGP and $n_b=99,150$}   \\
    OLS&--0.015&0.009&&--0.017&0.008&&0.\phantom{000}&0.009 \tabularnewline
Lasso with CV (2nd order poly)&0.089&--0.806&&0.100&--0.802&&--0.026&--0.721 \tabularnewline
Ridge with CV (2nd order poly)&--0.171&0.792&&--0.177&0.789&&--0.053&0.696 \tabularnewline
Lasso with CV (10th order poly)&0.285&--0.247&&0.229&--0.243&&0.358&--0.325 \tabularnewline
Ridge with CV (10th order poly)&--0.338&0.320&&--0.282&0.315&&--0.425&0.383 \tabularnewline
Random forest (low regularization)&0.142&--0.011&&0.142&--0.011&&0.168&--0.013 \tabularnewline
Random forest (high regularization)&--0.111&0.069&&--0.108&0.070&&--0.125&0.079 \tabularnewline
Gradient boosting (low regularization)&2.368&1.233&&2.356&1.236&&2.353&1.332 \tabularnewline
Gradient boosting (high regularization)&--1.443&--0.403&&--1.435&--0.406&&--1.424&--0.498 \tabularnewline
Neural net&--45.762&0.030&&0.080&0.033&&0.066&0.040 \tabularnewline
\\[-.3cm]
    \midrule
    \bottomrule
    \end{tabular}\par\medskip
    \parbox{\linewidth}{\textit{Notes:} The table shows the (average) stacking weights of each candidate learner for conventional stacking, pooled stacking and short-stacking using OLS as the final learner from the simulation example in Section \ref{sec:advantages_stacking}. The bootstrap sample size is denoted by $n_b$. Results are based on $1\,000$ replications. See Table~\ref{tab:sim_advantages_bias} for more information.
    }
\end{table}

\begin{table}[htbp]
    \scriptsize\singlespacing\centering
    \caption{Average stacking weights using single-best}
    \label{tab:sim_advantages_weights_singlebest}
    \begin{tabular}{lccccccccc}
    \toprule
    \midrule
        & \multicolumn{2}{c}{Stacking}&       & \multicolumn{2}{c}{Pooled stacking} &       & \multicolumn{2}{c}{Short-stacking} \\
        \cline{2-3} \cline{5-6} \cline{8-9}
                 & $E[Y\vert X]$ & $E[D\vert X]$ &     & $E[Y\vert X]$ & $E[D\vert X]$&     & $E[Y\vert X]$ & $E[D\vert X]$ \\
    \midrule
    \textit{Panel (A): Linear DGP and $n_b=9,915$}    \\
    OLS&0.812&0.645&&0.889&0.750&&0.821&0.650 \tabularnewline
Lasso with CV (2nd order poly)&0.163&0.274&&0.108&0.234&&0.150&0.266 \tabularnewline
Ridge with CV (2nd order poly)&0.021&0.016&&0.003&0.003&&0.028&0.027 \tabularnewline
Lasso with CV (10th order poly)&0.002&0.048&&0.\phantom{000}&0.011&&0.001&0.034 \tabularnewline
Ridge with CV (10th order poly)&0.002&0.010&&0.\phantom{000}&0.001&&0.\phantom{000}&0.023 \tabularnewline
Random forest (low regularization)&0.\phantom{000}&0.\phantom{000}&&0.\phantom{000}&0.\phantom{000}&&0.\phantom{000}&0.\phantom{000} \tabularnewline
Random forest (high regularization)&0.\phantom{000}&0.\phantom{000}&&0.\phantom{000}&0.\phantom{000}&&0.\phantom{000}&0.\phantom{000} \tabularnewline
Gradient boosting (low regularization)&0.\phantom{000}&0.003&&0.\phantom{000}&0.001&&0.\phantom{000}&0.\phantom{000} \tabularnewline
Gradient boosting (high regularization)&0.\phantom{000}&0.003&&0.\phantom{000}&0.\phantom{000}&&0.\phantom{000}&0.\phantom{000} \tabularnewline
Neural net&0.\phantom{000}&0.\phantom{000}&&0.\phantom{000}&0.\phantom{000}&&0.\phantom{000}&0.\phantom{000} \tabularnewline
 \\
    \textit{Panel (B): Linear DGP and $n_b=99,150$}    \\
    OLS&0.895&0.292&&0.960&0.288&&0.880&0.181 \tabularnewline
Lasso with CV (2nd order poly)&0.087&0.093&&0.037&0.073&&0.078&0.059 \tabularnewline
Ridge with CV (2nd order poly)&0.018&0.016&&0.003&0.003&&0.042&0.017 \tabularnewline
Lasso with CV (10th order poly)&0.\phantom{000}&0.186&&0.\phantom{000}&0.172&&0.\phantom{000}&0.151 \tabularnewline
Ridge with CV (10th order poly)&0.\phantom{000}&0.410&&0.\phantom{000}&0.464&&0.\phantom{000}&0.591 \tabularnewline
Random forest (low regularization)&0.\phantom{000}&0.\phantom{000}&&0.\phantom{000}&0.\phantom{000}&&0.\phantom{000}&0.\phantom{000} \tabularnewline
Random forest (high regularization)&0.\phantom{000}&0.\phantom{000}&&0.\phantom{000}&0.\phantom{000}&&0.\phantom{000}&0.\phantom{000} \tabularnewline
Gradient boosting (low regularization)&0.\phantom{000}&0.002&&0.\phantom{000}&0.\phantom{000}&&0.\phantom{000}&0.001 \tabularnewline
Gradient boosting (high regularization)&0.\phantom{000}&0.\phantom{000}&&0.\phantom{000}&0.\phantom{000}&&0.\phantom{000}&0.\phantom{000} \tabularnewline
Neural net&0.\phantom{000}&0.\phantom{000}&&0.\phantom{000}&0.\phantom{000}&&0.\phantom{000}&0.\phantom{000} \tabularnewline

          &       &       &       &       &  \\
    \textit{Panel (C): Non-linear DGP and $n_b=9,915$}    \\
    OLS&0.\phantom{000}&0.\phantom{000}&&0.\phantom{000}&0.\phantom{000}&&0.\phantom{000}&0.\phantom{000} \tabularnewline
Lasso with CV (2nd order poly)&0.093&0.149&&0.066&0.144&&0.054&0.081 \tabularnewline
Ridge with CV (2nd order poly)&0.132&0.124&&0.126&0.106&&0.058&0.093 \tabularnewline
Lasso with CV (10th order poly)&0.072&0.058&&0.040&0.028&&0.039&0.026 \tabularnewline
Ridge with CV (10th order poly)&0.019&0.041&&0.007&0.015&&0.005&0.037 \tabularnewline
Random forest (low regularization)&0.\phantom{000}&0.\phantom{000}&&0.\phantom{000}&0.\phantom{000}&&0.\phantom{000}&0.\phantom{000} \tabularnewline
Random forest (high regularization)&0.002&0.001&&0.\phantom{000}&0.\phantom{000}&&0.002&0.\phantom{000} \tabularnewline
Gradient boosting (low regularization)&0.673&0.357&&0.759&0.403&&0.832&0.623 \tabularnewline
Gradient boosting (high regularization)&0.009&0.268&&0.002&0.304&&0.008&0.140 \tabularnewline
Neural net&0.002&0.002&&0.\phantom{000}&0.\phantom{000}&&0.002&0.\phantom{000} \tabularnewline
 \\  
    \textit{Panel (D): Non-linear DGP and $n_b=99,150$}    \\
    OLS&0.\phantom{000}&0.\phantom{000}&&0.\phantom{000}&0.\phantom{000}&&0.\phantom{000}&0.\phantom{000} \tabularnewline
Lasso with CV (2nd order poly)&0.\phantom{000}&0.\phantom{000}&&0.\phantom{000}&0.\phantom{000}&&0.\phantom{000}&0.\phantom{000} \tabularnewline
Ridge with CV (2nd order poly)&0.\phantom{000}&0.\phantom{000}&&0.\phantom{000}&0.\phantom{000}&&0.\phantom{000}&0.\phantom{000} \tabularnewline
Lasso with CV (10th order poly)&0.\phantom{000}&0.\phantom{000}&&0.\phantom{000}&0.\phantom{000}&&0.\phantom{000}&0.\phantom{000} \tabularnewline
Ridge with CV (10th order poly)&0.\phantom{000}&0.\phantom{000}&&0.\phantom{000}&0.\phantom{000}&&0.\phantom{000}&0.\phantom{000} \tabularnewline
Random forest (low regularization)&0.\phantom{000}&0.\phantom{000}&&0.\phantom{000}&0.\phantom{000}&&0.\phantom{000}&0.\phantom{000} \tabularnewline
Random forest (high regularization)&0.\phantom{000}&0.\phantom{000}&&0.\phantom{000}&0.\phantom{000}&&0.\phantom{000}&0.\phantom{000} \tabularnewline
Gradient boosting (low regularization)&1.\phantom{000}&1.\phantom{000}&&1.\phantom{000}&1.\phantom{000}&&1.\phantom{000}&1.\phantom{000} \tabularnewline
Gradient boosting (high regularization)&0.\phantom{000}&0.\phantom{000}&&0.\phantom{000}&0.\phantom{000}&&0.\phantom{000}&0.\phantom{000} \tabularnewline
Neural net&0.\phantom{000}&0.\phantom{000}&&0.\phantom{000}&0.\phantom{000}&&0.\phantom{000}&0.\phantom{000} \tabularnewline
\\[-.3cm]
    \midrule
    \bottomrule
    \end{tabular}\par\medskip
        \parbox{\linewidth}{{\it Notes:} The table shows the (average) rates at which each candidate learner is selected by the single-best final learner when using conventional stacking, pooled stacking and short-stacking. 
        The bootstrap sample size is denoted by $n_b$. Results are based on $1\,000$ replications. See Table~\ref{tab:sim_advantages_bias} for more information.
       }
\end{table}

\begin{table}[htbp]
        \centering\scriptsize
        \caption{Computational time of DDML with conventional and short-stacking}
        \label{tab:timing}
        \begin{tabular}{rrrrrrrrrr}
        \toprule
        \midrule
        \multirow{3}{*}{Folds $K$} & \multirow{3}{*}{Obs.} & & \multicolumn{2}{c}{DDML} &  & & \\
        \cline{4-5}
        & & & \multicolumn{2}{c}{Stacking} & \multirow{2}{*}{OLS} & \multicolumn{1}{c}{PDS} & \multirow{2}{*}{Ratio} \\
        & & & \multicolumn{1}{c}{Conv.} & \multicolumn{1}{c}{Short} & & \multicolumn{1}{c}{lasso} & \\
        \midrule
             2&200&&14.31&3.81&0.0045&0.0461&0.2661 \tabularnewline
&400&&14.93&4.05&0.0047&0.0492&0.2711 \tabularnewline
&800&&18.36&4.95&0.0050&0.0518&0.2696 \tabularnewline
&1600&&26.82&7.07&0.0056&0.0595&0.2636 \tabularnewline
&9915&&138.39&34.35&0.0116&0.1488&0.2482 \tabularnewline
&99150&&2687.07&620.98&0.1013&1.4431&0.2311 \tabularnewline
5&200&&35.77&8.28&0.0045&0.0573&0.2315 \tabularnewline
&400&&41.83&9.79&0.0047&0.0589&0.2342 \tabularnewline
&800&&56.38&13.47&0.0049&0.0624&0.2388 \tabularnewline
&1600&&91.76&21.34&0.0056&0.0711&0.2326 \tabularnewline
&9915&&589.14&136.35&0.0110&0.1508&0.2314 \tabularnewline
10&200&&72.79&16.19&0.0046&0.0423&0.2224 \tabularnewline
&400&&85.87&19.27&0.0046&0.0524&0.2244 \tabularnewline
&800&&119.48&27.80&0.0049&0.0468&0.2327 \tabularnewline
&1600&&197.85&45.59&0.0054&0.0618&0.2304 \tabularnewline
&9915&&1364.07&313.84&0.0113&0.1426&0.2301 \tabularnewline
 \\[-.3cm]
        \midrule
        \bottomrule
        \end{tabular}\par\medskip
        \parbox{.58\linewidth}{{\it Notes:} The table reports the computational time in seconds of DDML paired with conventional stacking (`Conv.') or short-stacking (`Short') as implemented in \citet{Ahrens2023_ddml}, OLS as implemented in Stata's \texttt{regress}, post-double-selection lasso as implemented in \texttt{pdslasso} \citep{Ahrens2018}. DDML uses $V=5$ cross-validation folds and $K$ cross-fitting folds as indicated. Times reported are in seconds (average over 1\,000 replications). The computations were performed on the high-performance cluster of the ETH Zurich. Each instance used a single core of an AMD EPYC processor with 2.25-2.6GHz (nominal)/3.3-3.5 GHz (peak) and 4GB RAM.}
\end{table}

\clearpage\section{DDML and stacking in very small samples}
\setcounter{figure}{0}
\setcounter{table}{0}

\begin{table}[htbp]
        \centering\scriptsize
        \caption{Estimates based on the full sample ($N=9\,915$).}
        \label{tab:WZ_fullsample}
        \begin{tabular}{llcc}
        \hline\hline
        \multicolumn{2}{l}{\it Estimator} & \it Estimate   \\
        \hline
        \multicolumn{3}{l}{\it Panel A. No sample splitting}\\
        &OLS TWI&6751.907 \tabularnewline
&OLS QSI&5988.413 \tabularnewline
&Post double Lasso TWI c=0.5&6562.923 \tabularnewline
&Post double Lasso QSI c=0.5&5648.14 \tabularnewline
&Post double Lasso TWI c=1&6630.751 \tabularnewline
&Post double Lasso QSI c=1&4646.575 \tabularnewline
&Post double Lasso TWI c=1.5&7474.508 \tabularnewline
&Post double Lasso QSI c=1.5&4472.324 \tabularnewline
 
        \\
        \multicolumn{3}{l}{\it Panel B. DDML with candidate learners}\\
        &Neural net&6433.092 \tabularnewline
&OLS&6463.73 \tabularnewline
&Lasso with CV (TWI)&6780.161 \tabularnewline
&Ridge with CV (TWI)&6760.134 \tabularnewline
&Lasso with CV (QSI)&5722.624 \tabularnewline
&Ridge with CV (QSI)&5995.346 \tabularnewline
&Random forest (low regularization)&6089.389 \tabularnewline
&Random forest (high regularization)&6552.221 \tabularnewline
&Gradient boosting (low regularization)&7003.373 \tabularnewline
&Gradient boosting (high regularization)&7992.538 \tabularnewline

        \\
        \multicolumn{3}{l}{\it Panel C. DDML with stacking approaches}\\
        &Neural net&6433.092 \tabularnewline
&OLS&6463.73 \tabularnewline
&Lasso with CV (TWI)&6780.161 \tabularnewline
&Ridge with CV (TWI)&6760.134 \tabularnewline
&Lasso with CV (QSI)&5722.624 \tabularnewline
&Ridge with CV (QSI)&5995.346 \tabularnewline
&Random forest (low regularization)&6089.389 \tabularnewline
&Random forest (high regularization)&6552.221 \tabularnewline
&Gradient boosting (low regularization)&7003.373 \tabularnewline
&Gradient boosting (high regularization)&7992.538 \tabularnewline
\\[-.3cm]
        \hline\hline
        \end{tabular}\par\medskip
            \parbox{.7\linewidth}{{\it Notes:}   In the case of DDML estimators, the average estimates and standard errors are based on 50 replications. Panel A is reproduced from Table~1 in WZ. }
\end{table}

\begin{table}[htbp]
        \centering\scriptsize
        \caption{Short-stacking weights using CLS}
        \label{tab:WZ_ssw_weights}
        \begin{tabular}{llrrrrrrr}\hline\hline
        \multicolumn{2}{l}{\it Estimator} & \multicolumn{7}{c}{\it Observations}\\
        && 200 & 400 & 600 & 800 & 1\,200 & 1\,600 & 9\,915 \\ \hline
        \multicolumn{4}{l}{\it Panel A. $E[Y|X]$, $K=10$}\\
        \midrule
        &OLS&.164&.152&.115&.079&.037&.019&0 \tabularnewline
&Neural net&.047&.045&.048&.067&.098&.05&.076 \tabularnewline
&Lasso with CV (TWI)&.043&.034&.034&.035&.03&.033&.091 \tabularnewline
&Ridge with CV (TWI)&.056&.048&.041&.025&.011&.006&.032 \tabularnewline
&Lasso with CV (QSI)&.252&.274&.266&.264&.271&.297&.639 \tabularnewline
&Ridge with CV (QSI)&.194&.252&.297&.328&.341&.357&.153 \tabularnewline
&Random forest (low regularization)&.095&.097&.113&.131&.161&.2&.01 \tabularnewline
&Random forest (high regularization)&.081&.04&.025&.021&.018&.016&0 \tabularnewline
&Gradient boosting (low regularization)&.041&.04&.049&.041&.03&.021&0 \tabularnewline
&Gradient boosting (high regularization)&.028&.019&.013&.009&.002&.001&0 \tabularnewline
\\
        \multicolumn{4}{l}{\it Panel B. $E[D|X]$, $K=10$}\\
        \midrule
        &OLS&.132&.196&.234&.252&.245&.257&.163 \tabularnewline
&Neural net&.04&.041&.038&.036&.031&.029&.038 \tabularnewline
&Lasso with CV (TWI)&.053&.031&.025&.02&.016&.012&.106 \tabularnewline
&Ridge with CV (TWI)&.038&.018&.013&.015&.008&.005&.029 \tabularnewline
&Lasso with CV (QSI)&.173&.225&.25&.248&.25&.228&.413 \tabularnewline
&Ridge with CV (QSI)&.202&.124&.072&.06&.068&.064&0 \tabularnewline
&Random forest (low regularization)&.103&.123&.144&.187&.249&.307&.006 \tabularnewline
&Random forest (high regularization)&.159&.129&.107&.09&.051&.031&.102 \tabularnewline
&Gradient boosting (low regularization)&.043&.046&.054&.047&.045&.041&.144 \tabularnewline
&Gradient boosting (high regularization)&.059&.065&.064&.046&.038&.025&0 \tabularnewline
\\
        \multicolumn{4}{l}{\it Panel C. $E[Y|X]$, $K=10$}\\
        \midrule
        &OLS&.122&.098&.066&.026&.003&.001&0 \tabularnewline
&Neural net&0&0&0&0&0&0&0 \tabularnewline
&Lasso with CV (TWI)&.03&.022&.01&.013&.014&.023&0 \tabularnewline
&Ridge with CV (TWI)&.074&.077&.079&.052&.03&.013&0 \tabularnewline
&Lasso with CV (QSI)&.323&.376&.361&.381&.393&.405&.995 \tabularnewline
&Ridge with CV (QSI)&.239&.314&.379&.428&.478&.479&.005 \tabularnewline
&Random forest (low regularization)&.129&.058&.05&.049&.049&.044&0 \tabularnewline
&Random forest (high regularization)&.022&.005&.001&.001&0&.001&0 \tabularnewline
&Gradient boosting (low regularization)&.025&.033&.046&.046&.032&.034&0 \tabularnewline
&Gradient boosting (high regularization)&.035&.016&.009&.004&0&0&0 \tabularnewline
\\
        \multicolumn{4}{l}{\it Panel D. $E[D|X]$, $K=10$}\\
        \midrule
        &OLS&.038&.108&.17&.189&.173&.132&.005 \tabularnewline
&Neural net&0&0&0&0&0&0&0 \tabularnewline
&Lasso with CV (TWI)&.058&.032&.017&.011&.005&.003&.002 \tabularnewline
&Ridge with CV (TWI)&.06&.013&.009&.01&.002&.001&0 \tabularnewline
&Lasso with CV (QSI)&.232&.309&.313&.287&.261&.168&.754 \tabularnewline
&Ridge with CV (QSI)&.242&.105&.032&.034&.05&.032&0 \tabularnewline
&Random forest (low regularization)&.079&.028&.011&.008&.004&.003&0 \tabularnewline
&Random forest (high regularization)&.185&.249&.256&.304&.344&.507&0 \tabularnewline
&Gradient boosting (low regularization)&.009&.022&.048&.064&.115&.141&.24 \tabularnewline
&Gradient boosting (high regularization)&.098&.135&.143&.092&.046&.013&0 \tabularnewline
\\[-.3cm]
        \hline\hline
        \end{tabular}\par\medskip
        \parbox{\linewidth}{{\it Notes:}   The table reports the stacking weights corresponding to the DDML stacking estimator in Figure~\ref{tab:WZ_bias}. The stacking weights are averaged over folds, based on 10-fold cross-fitting and shows for the estimation of $E[Y|X]$ and $E[D|X]$ in Panel~A and B, respectively. See notes below Table~\ref{tab:WZ_bias} for more information.}
\end{table}

\begin{sidewaystable} 
        \centering\scriptsize\singlespacing
        \caption{Mean bias under linear DGP in small samples based on the calibrated Monte Carlo in Section~\ref{sec:advantages_stacking} }\label{tab:small_sample_bias_dgp0}
        \sisetup{
        round-mode=places,
        round-precision=1,
        tight-spacing=true,
        table-format=-4.1,
        table-number-alignment=center
        }
        \label{tab:sim_advantages_bias_small_linear}
        \begin{tabular}{rlr@{\hskip 0.1in}rr@{\hskip 0.1in}rr@{\hskip 0.1in}rr@{\hskip 0.1in}rr@{\hskip 0.1in}r}
        \toprule
        \midrule
            & & \multicolumn{10}{c}{\it Bootstrap sample size $n_b$} \\
        \multicolumn{2}{l}{}& \multicolumn{2}{c}{200}  & \multicolumn{2}{c}{400}    & \multicolumn{2}{c}{800}  &  \multicolumn{2}{c}{1\,600}     & \multicolumn{2}{c}{9\,915}     \\
        \cmidrule{3-12} 
        &&\multicolumn{1}{c}{Bias} & \multicolumn{1}{c}{s.e.}  &\multicolumn{1}{c}{Bias} & \multicolumn{1}{c}{s.e.}&\multicolumn{1}{c}{Bias} & \multicolumn{1}{c}{s.e.}&\multicolumn{1}{c}{Bias} & \multicolumn{1}{c}{s.e.}&\multicolumn{1}{c}{Bias} & \multicolumn{1}{c}{s.e.}\\
        \midrule
              \multicolumn{4}{l}{Full sample estimators:}\\
              \partialinput{1}{2}{Simul/sim_Advantages/sim_output_dgp0_10folds_withse}
               \multicolumn{4}{l}{DDML methods:}\\
               \multicolumn{4}{l}{\it ~~Candidate learners ($K=10$)}\\
              \partialinput{3}{12}{Simul/sim_Advantages/sim_output_dgp0_10folds_withse}
              \multicolumn{4}{l}{\it ~~Meta learners ($K=10$)}\\
              \partialinput{13}{24}{Simul/sim_Advantages/sim_output_dgp0_10folds_withse}
              \multicolumn{4}{l}{\it ~~Meta learners ($K=2$)}\\
              \partialinput{13}{24}{Simul/sim_Advantages/sim_output_dgp0_2folds_withse}\\[-.3cm]
        \midrule
        \bottomrule
        \end{tabular}
        \par\medskip
      \parbox{\linewidth}{%
      \textit{Notes:} The table reports mean bias and associated standard errors for the listed estimators. 
      We consider DDML with the following individual learners: OLS with elementary covariates, CV lasso and CV ridge with second-order polynomials and interactions, CV lasso and CV ridge with 10th-order polynomials but no interactions, random forest with low regularization (8 predictors considered at each leaf split, no limit on the number of observations per node, bootstrap sample size of 70\%), highly regularized random forest (5 predictors considered at each leaf split, at least 10 observation per node, bootstrap sample size of 70\%), gradient-boosted trees with low regularization (500 trees, maximum depth of 3 and a learning rate of 0.01), gradient-boosted trees with high regularization: 250 trees, maximum depth of 3 and a learning rate of 0.01, feed-forward neural nets with three hidden layers of size five.
      For reference, we report two estimators using the full sample: OLS and PDS lasso.  We report results for four meta learners: Stacking with CLS, short-stacking with CLS, single best overall and single best by fold. 
      Results are based on 1\,000 replications.
      }
\end{sidewaystable}

\begin{sidewaystable} 
        \centering\scriptsize\singlespacing
        \caption{Mean bias under non-linear DGP in small samples based on the calibrated Monte Carlo in Section~\ref{sec:advantages_stacking}}\label{tab:small_sample_bias_dgp1}
        \sisetup{
        round-mode=places,
        round-precision=1,
        tight-spacing=true,
        table-format=-4.1,
        table-number-alignment=center
        }
        \label{tab:sim_advantages_bias_small_nonlinear}
        \begin{tabular}{rlr@{\hskip 0.1in}rr@{\hskip 0.1in}rr@{\hskip 0.1in}rr@{\hskip 0.1in}rr@{\hskip 0.1in}r}
        \toprule
        \midrule
            & & \multicolumn{10}{c}{\it Bootstrap sample size $n_b$} \\
        \multicolumn{2}{l}{}& \multicolumn{2}{c}{200}  & \multicolumn{2}{c}{400}    & \multicolumn{2}{c}{800}  &  \multicolumn{2}{c}{1\,600}     & \multicolumn{2}{c}{9\,915}     \\
        \cmidrule{3-12} 
        &&\multicolumn{1}{c}{Bias} & \multicolumn{1}{c}{s.e.}  &\multicolumn{1}{c}{Bias} & \multicolumn{1}{c}{s.e.}&\multicolumn{1}{c}{Bias} & \multicolumn{1}{c}{s.e.}&\multicolumn{1}{c}{Bias} & \multicolumn{1}{c}{s.e.}&\multicolumn{1}{c}{Bias} & \multicolumn{1}{c}{s.e.}\\
        \midrule
              \multicolumn{4}{l}{Full sample estimators:}\\
              \partialinput{1}{2}{Simul/sim_Advantages/sim_output_dgp1_10folds_withse}
               \multicolumn{4}{l}{DDML methods:}\\
               \multicolumn{4}{l}{\it ~~Candidate learners ($K=10$)}\\
              \partialinput{3}{12}{Simul/sim_Advantages/sim_output_dgp1_10folds_withse}
              \multicolumn{4}{l}{\it ~~Meta learners ($K=10$)}\\
              \partialinput{13}{24}{Simul/sim_Advantages/sim_output_dgp1_10folds_withse}
              \multicolumn{4}{l}{\it ~~Meta learners ($K=2$)}\\
              \partialinput{13}{24}{Simul/sim_Advantages/sim_output_dgp1_2folds_withse}\\[-.3cm]
        \midrule
        \bottomrule
        \end{tabular}
        \par\medskip
      \parbox{.8\linewidth}{%
      \textit{Notes:} See Table~\ref{tab:small_sample_bias_dgp0} notes.
      }
\end{sidewaystable}

\begin{sidewaystable}
        \centering\scriptsize\singlespacing
        \caption{Coverage in small samples based on the calibrated Monte Carlo in Section~\ref{sec:advantages_stacking}}\label{tab:small_sample_cov}
        \sisetup{
        round-mode=places,
        round-precision=2,
        tight-spacing=true,
        table-format=-1.2,
        table-number-alignment=center 
        }
        \label{tab:sim_advantages_coverage_small}
        \begin{tabular}{rlSSSSScSSSSS}
        \toprule
        \midrule
            & & \multicolumn{11}{c}{\it Bootstrap sample size $n_b$} \\
           & & \multicolumn{5}{c}{\it Panel A. Linear DGP} &&\multicolumn{5}{c}{\it Panel B. Non-linear DGP} \\
        \cmidrule{3-7} \cmidrule{9-13} 
        \multicolumn{2}{l}{}& \multicolumn{1}{c}{200}  & \multicolumn{1}{c}{400}    & \multicolumn{1}{c}{800}  &  \multicolumn{1}{c}{1600}     & \multicolumn{1}{c}{9915}  && \multicolumn{1}{c}{200}  & \multicolumn{1}{c}{400}    & \multicolumn{1}{c}{800}  &  \multicolumn{1}{c}{1\,600}     & \multicolumn{1}{c}{9\,915}   \\ \midrule
              \multicolumn{4}{l}{Full sample estimators:}\\
              \partialinput{1}{2}{Simul/sim_Advantages/sim_small_cover_wide_folds10}
               \multicolumn{4}{l}{DDML methods:}\\
               \multicolumn{4}{l}{\it ~~Candidate learners ($K=10$)}\\
              \partialinput{3}{12}{Simul/sim_Advantages/sim_small_cover_wide_folds10}
              \multicolumn{4}{l}{\it ~~Meta learners ($K=10$)}\\
              \partialinput{13}{24}{Simul/sim_Advantages/sim_small_cover_wide_folds10}
              \multicolumn{4}{l}{\it ~~Meta learners ($K=2$)}\\
              \partialinput{13}{13}{Simul/sim_Advantages/sim_small_cover_wide_folds2}
              \partialinput{16}{24}{Simul/sim_Advantages/sim_small_cover_wide_folds2} \\[-.3cm]
        \midrule
        \bottomrule
        \end{tabular}
        \par\medskip
      \parbox{.8\linewidth}{
      \textit{Notes:} This table reports coverage of 95\% interval estimates in the small sample simulation. See Table~\ref{tab:small_sample_bias_dgp0} notes for more detail.
      }  
\end{sidewaystable}

\section{Gender citation gap}
\setcounter{figure}{0}
\setcounter{table}{0}

\begin{figure}[htbp]
        \centering\singlespacing\scriptsize
        \caption{The citation gap by authors' gender composition}
        \label{fig:scopus_cites_otherthresholds}
        \begin{subfigure}{\linewidth}
            \includegraphics[width=\linewidth]{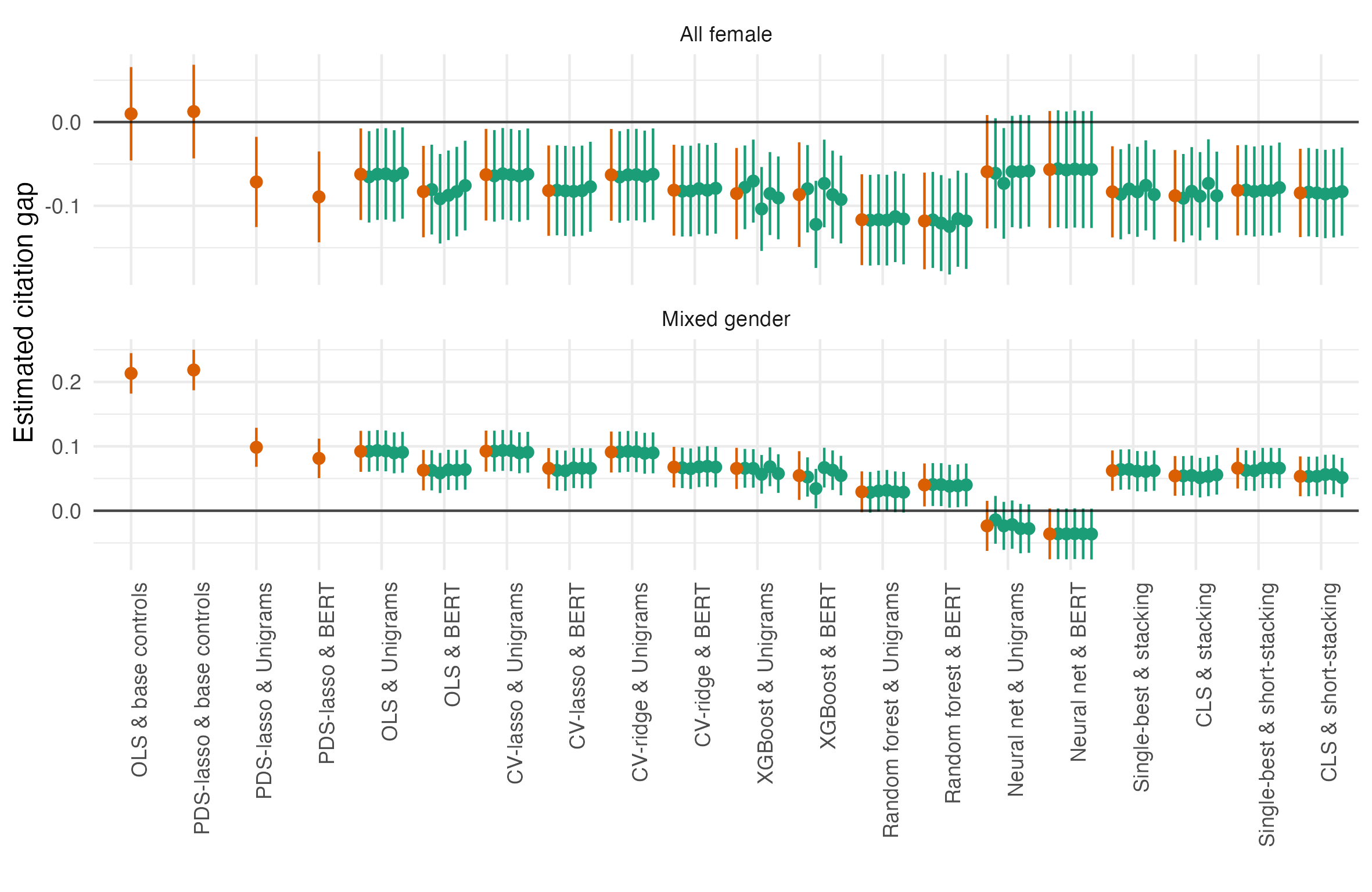}\\[-.5cm]
        \parbox{\linewidth}{\hspace*{1.4cm}$\underbracket{\hspace{2.8cm}}_{\textsf{Full-sample estimators}}$~~$\underbracket{\hspace{8.5cm}}_{\textsf{DDML with candidate learners}}$~~$\underbracket{\hspace{2.6cm}}_{\textsf{DDML and stacking}}$}\par
             \caption{Threshold 60\%}
        \end{subfigure}
        \begin{subfigure}{\linewidth}
            \includegraphics[width=\linewidth]{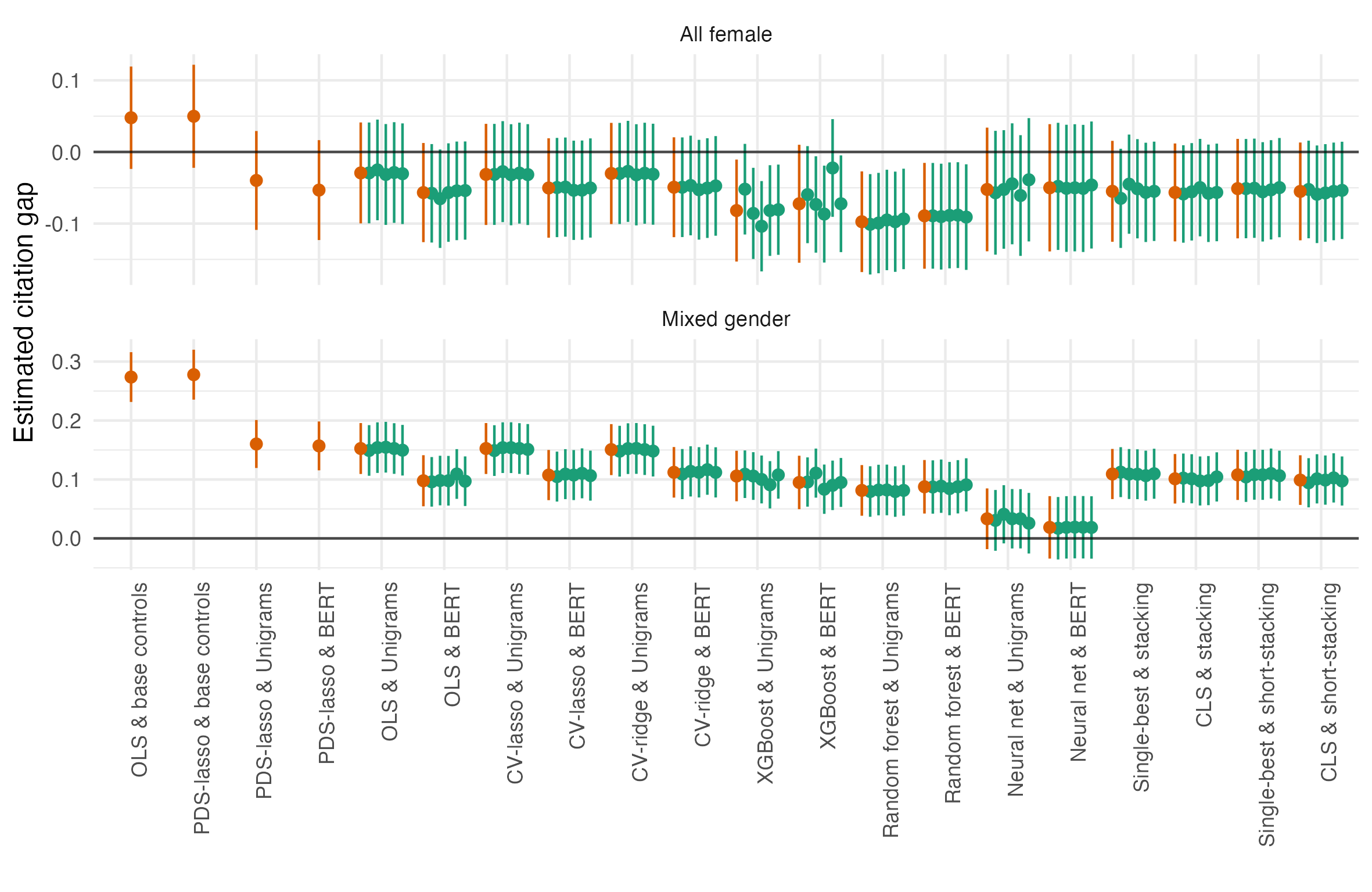}\\[-.5cm]
        \parbox{\linewidth}{\hspace*{1.4cm}$\underbracket{\hspace{2.8cm}}_{\textsf{Full-sample estimators}}$~~$\underbracket{\hspace{8.5cm}}_{\textsf{DDML with candidate learners}}$~~$\underbracket{\hspace{2.6cm}}_{\textsf{DDML and stacking}}$}\par
            \caption{Threshold 90\%}
        \end{subfigure}
        \vspace*{.5cm}
        \parbox{\linewidth}{{\it Notes:}  
          The figure shows estimates of $\theta_0$ summarizing average relative difference in total citations between all-male and all-female authorship, and all-male and mixed-gender authorship, respectively using different thresholds for successful classification of an author's sex. See Figure~\ref{fig:scopus_cites_log} notes for more information.
        } 
\end{figure}

\begin{table}[htbp]
        \centering\scriptsize\singlespacing
        \caption{Estimates for the citation penalty of all-female and mixed-gender authored articles}
        \label{tab:scopus_cites_appendix}
        \begin{tabular}{lcccc}\toprule\toprule
        & \multicolumn{2}{c}{Log citations} & \multicolumn{2}{c}{Citation counts} \\ 
         & \it All female & \it Mixed gender & \it All female & \it Mixed gender \\
        \midrule
        \multicolumn{5}{l}{\it Panel A. Full-sample estimators}\\
        \partialinput{3}{8}{scopus_cites/results_tab}\\[-.3cm]
        \\
        \multicolumn{5}{l}{\it Panel B. DDML with candidate learners}\\
        \partialinput{9}{34}{scopus_cites/results_tab}\\[-.3cm]
        \\
        \multicolumn{5}{l}{\it Panel C. DDML with stacking approaches}\\
        \partialinput{35}{42}{scopus_cites/results_tab}\\[-.3cm]
        \bottomrule\bottomrule
        \end{tabular}\par\medskip
        \parbox{\linewidth}{\emph{Notes:} The table shows median-aggregated estimates of the gender citation gap for all--female and mixed-gender authored articles. We show results using both log citations and citation counts as the outcome variable. Standard errors are robust to heteroskedasticity.  
        See Figure~\ref{fig:scopus_cites_log} for information on the candidate learners and stacking approaches.}
\end{table}

\section{Gender wage gap}
\setcounter{figure}{0}
\setcounter{table}{0}

\begin{table}[htbp]
        \centering\scriptsize\singlespacing
        \caption{Stacking weights in the gender wage gap application.}
        \label{tab:gwg_weights}
        \resizebox{\linewidth}{!}{%
        \begin{tabular}{llccccccccccc}\toprule\toprule
        &\multicolumn{3}{c}{\it Conventional stacking} & &\multicolumn{3}{c}{\it Short-stacking} & & \multicolumn{3}{c}{\it Mean-squared error}  \\
        \cline{2-4} \cline{6-8} \cline{10-12}
        & $g_0(0,X)$  & $g_0(1,X)$  & $m_0(X)$ & & $g_0(0,X)$  & $g_0(1,X)$  & $m_0(X)$ && $g_0(0,X)$  & $g_0(1,X)$  & $m_0(X)$  \\
        \midrule
 OLS/logit & 0.023 & 0.012 & 0.242 &  & 0.027 & 0.013 & 0.211 &  & 0.369 & 0.347 & 0.161 \\ 
  OLS/logit (simple) & 0.004 & 0.\phantom{000} & 0.\phantom{000} &  & 0.\phantom{000} & 0.\phantom{000} & 0.\phantom{000} &  & 0.267 & 0.204 & 0.223 \\ 
  CV-lasso & 0.103 & 0.136 & 0.109 &  & 0.03\phantom{0} & 0.076 & 0.047 &  & 0.236 & 0.178 & 0.16\phantom{0} \\ 
  CV-ridge & 0.189 & 0.04\phantom{0} & 0.064 &  & 0.225 & 0.024 & 0.108 &  & 0.237 & 0.18\phantom{0} & 0.161 \\ 
  CV-lasso (extended) & 0.041 & 0.157 & 0.016 &  & 0.035 & 0.266 & 0.002 &  & 0.238 & 0.18\phantom{0} & 0.161 \\ 
  CV-ridge (extended) & 0.011 & 0.04\phantom{0} & 0.011 &  & 0.003 & 0.024 & 0.022 &  & 0.336 & 0.194 & 0.161 \\ 
  Random forest 1 & 0.435 & 0.506 & 0.275 &  & 0.483 & 0.507 & 0.28\phantom{0} &  & 0.23\phantom{0} & 0.176 & 0.161 \\ 
  Random forest 2 & 0.\phantom{000} & 0.\phantom{000} & 0.\phantom{000} &  & 0.\phantom{000} & 0.\phantom{000} & 0.\phantom{000} &  & 0.258 & 0.19\phantom{0} & 0.171 \\ 
  Random forest 3 & 0.\phantom{000} & 0.\phantom{000} & 0.\phantom{000} &  & 0.\phantom{000} & 0.\phantom{000} & 0.\phantom{000} &  & 0.274 & 0.199 & 0.179 \\ 
  Gradient boosting 1 & 0.025 & 0.008 & 0.039 &  & 0.011 & 0.003 & 0.022 &  & 0.239 & 0.183 & 0.16\phantom{0} \\ 
  Gradient boosting 2 & 0.15\phantom{0} & 0.059 & 0.216 &  & 0.175 & 0.063 & 0.285 &  & 0.254 & 0.196 & 0.161 \\ 
  Neural net 1 & 0.013 & 0.022 & 0.\phantom{000} &  & 0.\phantom{000} & 0.\phantom{000} & 0.\phantom{000} &  & 0.349 & 0.263 & 0.241 \\ 
  Neural net 2 & 0.008 & 0.02\phantom{0} & 0.027 &  & 0.01\phantom{0} & 0.023 & 0.023 &  & 0.643 & 0.357 & 0.176 \\ 
  \\[-.3cm]
        \bottomrule\bottomrule
        \end{tabular}}\par\medskip
        \parbox{\linewidth}{\emph{Notes:} The table shows weights of conventional and short-stacking along with the mean-squared prediction error by candidate learners and by variable. The final learner is constrained least squares. The stacking weights are averaged over cross-fitting repetitions. Pooled stacking weights are shown in Appendix Table~\ref{tab:gwg_weights_pooled}.}
\end{table}

\begin{table}[htbp]
        \scriptsize\singlespacing\centering
        \caption{Stacking weights of pooled stacking using constrained least squares.}
        \label{tab:gwg_weights_pooled}
        \begin{tabular}{lcccccccccccc}
        \toprule
        \midrule
            &  \multicolumn{3}{c}{\it Pooled stacking}  \\
            \cmidrule{2-4}  
         & $g_0(0,X)$  & $g_0(1,X)$  & $m_0(X)$  \\
        \midrule
 OLS/logit & 0.014 & 0.001 & 0.257 \\ 
  OLS/logit (simple) & 0.\phantom{000} & 0.\phantom{000} & 0.\phantom{000} \\ 
  CV-lasso & 0.15\phantom{0} & 0.23\phantom{0} & 0.136 \\ 
  CV-ridge & 0.205 & 0.064 & 0.063 \\ 
  CV-lasso (extended) & 0.\phantom{000} & 0.078 & 0.\phantom{000} \\ 
  CV-ridge (extended) & 0.\phantom{000} & 0.019 & 0.\phantom{000} \\ 
  Random forest 1 & 0.462 & 0.521 & 0.288 \\ 
  Random forest 2 & 0.\phantom{000} & 0.\phantom{000} & 0.\phantom{000} \\ 
  Random forest 3 & 0.\phantom{000} & 0.\phantom{000} & 0.\phantom{000} \\ 
  Gradient boosting 1 & 0.\phantom{000} & 0.\phantom{000} & 0.008 \\ 
  Gradient boosting 2 & 0.165 & 0.071 & 0.23\phantom{0} \\ 
  Neural net 1 & 0.\phantom{000} & 0.\phantom{000} & 0.\phantom{000} \\ 
  Neural net 2 & 0.004 & 0.016 & 0.018 \\ 
  \\[-.3cm]
        \midrule
        \bottomrule
        \end{tabular}\par\medskip\scriptsize 
            \parbox{.6\linewidth}{{\it Notes:} The table shows pooled stacking weights for each of the considered candidate learners. The final learner is constrained least squares. The stacking weights are averaged over cross-fitting repetitions.}
\end{table}

\begin{table}[htbp]
        \centering\scriptsize\singlespacing
        \caption{Stacking weights using single-best final learner.}
        \label{tab:gwg_weights_singlebest}
        \resizebox{\linewidth}{!}{%
        \begin{tabular}{llccccccccccc}\toprule\toprule
        &\multicolumn{3}{c}{\it Conventional stacking} & &\multicolumn{3}{c}{\it Short-stacking} & & \multicolumn{3}{c}{\it Pooled stacking}  \\
        \cline{2-4} \cline{6-8} \cline{10-12}
        & $g_0(0,X)$  & $g_0(1,X)$  & $m_0(X)$ & & $g_0(0,X)$  & $g_0(1,X)$  & $m_0(X)$ && $g_0(0,X)$  & $g_0(1,X)$  & $m_0(X)$  \\
        \midrule
 OLS/logit & 0.\phantom{000} & 0.\phantom{000} & 0.\phantom{000} &  & 0.\phantom{000} & 0.\phantom{000} & 0.\phantom{000} &  & 0.\phantom{000} & 0.\phantom{000} & 0.\phantom{000} \\ 
  OLS/logit (simple) & 0.\phantom{000} & 0.\phantom{000} & 0.\phantom{000} &  & 0.\phantom{000} & 0.\phantom{000} & 0.\phantom{000} &  & 0.\phantom{000} & 0.\phantom{000} & 0.\phantom{000} \\ 
  CV-lasso & 0.06\phantom{0} & 0.03\phantom{0} & 0.79\phantom{0} &  & 0.\phantom{000} & 0.\phantom{000} & 1.\phantom{000} &  & 0.\phantom{000} & 0.\phantom{000} & 0.4\phantom{00} \\ 
  CV-ridge & 0.07\phantom{0} & 0.\phantom{000} & 0.04\phantom{0} &  & 0.\phantom{000} & 0.\phantom{000} & 0.\phantom{000} &  & 0.\phantom{000} & 0.\phantom{000} & 0.1\phantom{00} \\ 
  CV-lasso (extended) & 0.02\phantom{0} & 0.06\phantom{0} & 0.03\phantom{0} &  & 0.\phantom{000} & 0.\phantom{000} & 0.\phantom{000} &  & 0.\phantom{000} & 0.\phantom{000} & 0.1\phantom{00} \\ 
  CV-ridge (extended) & 0.\phantom{000} & 0.\phantom{000} & 0.01\phantom{0} &  & 0.\phantom{000} & 0.\phantom{000} & 0.\phantom{000} &  & 0.\phantom{000} & 0.\phantom{000} & 0.\phantom{000} \\ 
  Random forest 1 & 0.85\phantom{0} & 0.91\phantom{0} & 0.02\phantom{0} &  & 1.\phantom{000} & 1.\phantom{000} & 0.\phantom{000} &  & 1.\phantom{000} & 1.\phantom{000} & 0.\phantom{000} \\ 
  Random forest 2 & 0.\phantom{000} & 0.\phantom{000} & 0.\phantom{000} &  & 0.\phantom{000} & 0.\phantom{000} & 0.\phantom{000} &  & 0.\phantom{000} & 0.\phantom{000} & 0.\phantom{000} \\ 
  Random forest 3 & 0.\phantom{000} & 0.\phantom{000} & 0.\phantom{000} &  & 0.\phantom{000} & 0.\phantom{000} & 0.\phantom{000} &  & 0.\phantom{000} & 0.\phantom{000} & 0.\phantom{000} \\ 
  Gradient boosting 1 & 0.\phantom{000} & 0.\phantom{000} & 0.11\phantom{0} &  & 0.\phantom{000} & 0.\phantom{000} & 0.\phantom{000} &  & 0.\phantom{000} & 0.\phantom{000} & 0.4\phantom{00} \\ 
  Gradient boosting 2 & 0.\phantom{000} & 0.\phantom{000} & 0.\phantom{000} &  & 0.\phantom{000} & 0.\phantom{000} & 0.\phantom{000} &  & 0.\phantom{000} & 0.\phantom{000} & 0.\phantom{000} \\ 
  Neural net 1 & 0.\phantom{000} & 0.\phantom{000} & 0.\phantom{000} &  & 0.\phantom{000} & 0.\phantom{000} & 0.\phantom{000} &  & 0.\phantom{000} & 0.\phantom{000} & 0.\phantom{000} \\ 
  Neural net 2 & 0.\phantom{000} & 0.\phantom{000} & 0.\phantom{000} &  & 0.\phantom{000} & 0.\phantom{000} & 0.\phantom{000} &  & 0.\phantom{000} & 0.\phantom{000} & 0.\phantom{000} \\ 
  \\[-.3cm]
        \bottomrule\bottomrule
        \end{tabular}}\par\medskip
        \parbox{\linewidth}{\emph{Notes:} The table shows weights of conventional stacking, short-stacking and pooled stacking by candidate learners and by conditional expectation function. The stacking weights are averaged over cross-fitting repetitions.}
\end{table}

\begin{table}[htbp]
        \scriptsize\singlespacing\centering
        \caption{Median aggregate estimates by stacking approach and by final learner}
        \label{tab:gwg_results_1}
        \begin{tabular}{lcccc}
        \toprule
        \midrule
         & \multicolumn{4}{c}{\it Final learner} \\
         \cmidrule{2-5}
            &{\it Unweighted}& \multirow{2}{*}{\it CLS}     & \multirow{2}{*}{\it OLS}  &   {\it Single-}     \\
              & {\it average}&       &   &         {\it best}\\
        \midrule
        \partialinput{4}{9}{GWG/regression_results_1} \\[-.3cm]
        \midrule
        \bottomrule
        \end{tabular}\par\medskip\scriptsize 
        \parbox{.54\linewidth}{{\it Notes:} The table reports median aggregate estimates by stacking type and final learner. See Figure~\ref{fig:gwg} for more information.}
\end{table}

\begin{table}[htbp]
        \scriptsize\singlespacing\centering
        \caption{Median aggregate estimates for each candidate learner}
        \label{tab:gwg_results_2}
        \begin{tabular}{lcccc}
        \toprule
        \midrule
            & {\it Gender wage}  \\
              &{\it gap}\\
        \midrule
        \partialinput{4}{29}{GWG/regression_results_2}\\[-.3cm]
        \midrule
        Observations & 4836 \\
        \midrule
        \bottomrule
        \end{tabular}\par\medskip\scriptsize 
        \parbox{.4\linewidth}{{\it Notes:} The table reports median aggregate estimates by candidate learner. See Figure~\ref{fig:gwg} for more information.}
\end{table}

\end{document}